\documentclass[pre,aps,superscriptaddress]{revtex4}

\pdfoutput=1

\usepackage{amsmath,amsfonts,amssymb,bm,hyperref,graphicx}
\usepackage{color}

\begin{document}

\title{Symmetries and Geometrical Properties of Dynamical Fluctuations in Molecular Dynamics}
\author{Robert L. Jack}
\affiliation{Department of Applied Mathematics and Theoretical Physics, University of Cambridge, Wilberforce Road, Cambridge CB3 0WA, United Kingdom}
\affiliation{Department of Chemistry, University of Cambridge, Lensfield Road, Cambridge CB2 1EW, United Kingdom}
\affiliation{Department of Physics, University of Bath, Bath BA2 7AY, United Kingdom}
\author{Marcus Kaiser}
\affiliation{Department of Mathematical Sciences, University of Bath, Bath BA2 7AY, United Kingdom}
\author{Johannes Zimmer}
\affiliation{Department of Mathematical Sciences, University of Bath, Bath BA2 7AY, United Kingdom}

\newcommand{\ee}{\mathrm{e}}
\newcommand{\dd}{\mathrm{d}}

\newcommand{\eps}{\varepsilon}

\newcommand{\Pc}{{\mathcal{P}}}
\newcommand{\Pcbar}{\overline{\mathcal{P}}}

\newcommand{\Fc}{\mathcal{F}}
\newcommand{\Fcbar}{\overline{\Fc}}

\newcommand{\wbar}{\overline{w}}
\newcommand{\pibar}{\overline{\pi}}
\newcommand{\Ubar}{\overline{U}}
\newcommand{\fbar}{\overline{f}}
\newcommand{\phibar}{\overline{\phi}}
\newcommand{\jbar}{\overline{\jmath}}

\newcommand{\Jc}{\mathcal{J}}
\newcommand{\Jbar}{\overline{J}}
\newcommand{\Jcbar}{\overline{\Jc}}

\newcommand{\Vc}{\mathcal{V}}

\newcommand{\WW}{\mathbb{W}}
\newcommand{\WWbar}{\overline{\mathbb{W}}}
\newcommand{\CC}{\mathcal{C}}

\newcommand{\TT}{\mathbb{T}}
\newcommand{\PP}{\mathbb{P}}

\newcommand{\fext}{f^{\rm ext}}

\newcommand{\ddiv}{\operatorname{div}}

\newcommand{\rlj}[1]{{\color{blue}#1}}

\begin{abstract}
We describe some general results that constrain the dynamical fluctuations that can occur in non-equilibrium steady states, with a
focus on molecular dynamics.  That is, we consider Hamiltonian systems, coupled to external heat baths, and driven out of
equilibrium by non-conservative forces.  We focus on the probabilities of rare events (large deviations).  First, we discuss a PT
(parity-time) symmetry that appears in ensembles of trajectories where a current is constrained to have a large (non-typical)
value.  We analyse the heat flow in such ensembles, and compare it with non-equilibrium steady states.  Second, we consider
pathwise large deviations that are defined by considering many copies of a system.  We show how the probability currents in such
systems can be decomposed into orthogonal contributions, that are related to convergence to equilibrium and to dissipation.  We
discuss the implications of these results for modelling non-equilibrium steady states.
\end{abstract}

\maketitle

\section{Introduction}

This article studies dynamical fluctuations in stochastic processes of relevance for molecular dynamics. More precisely, we consider stochastic systems described by underdamped Langevin equations. We focus on large
deviation principles, which encode the probability of rare dynamical events~\cite{Touchette2009} and discuss the physical
principles and symmetries that govern the probabilities of such events. The applications we have in mind are physical systems of
interacting atoms and molecules, which are usually thought of as evolving by deterministic (Hamiltonian) dynamics. However, it is
now standard to add stochastic terms to these equations of motion to describe the coupling of these systems to their
environments. This coupling is especially important if we aim to describe non-equilibrium steady states, in which the work done
by external forces must be dissipated in the environment. For~this reason, a clear understanding of the interplay between
molecular dynamics and stochastic forces is vital in order to build accurate models of molecular systems away from equilibrium.

\subsection{Motivation}

Molecular dynamics~\cite{frenkelsmit} is now established as a standard tool for computational studies of a variety of systems,
including a wide range of biomolecules and physical materials. For a system that is completely isolated from its environment, the
prescription for computation of dynamical trajectories is extremely simple: one identifies a set of co-ordinates $q$, their
canonical momenta $p$, and a Hamiltonian $H$. The equations of motion are simply
\begin{equation} 
 \partial_t{q} = \partial H / \partial p, \qquad \partial_t{p} = -\partial H / \partial q . 
\end{equation}
Moreover, there are efficient computational methods for obtaining accurate solutions to these equations, which perform well on
modern high-performance computing platforms.

However, many physical systems are not completely isolated from their environments. In~particular, they often exchange energy with
some kind of thermal bath, so that they equilibrate at some temperature $T$. The volume of some systems may also fluctuate, so
that their pressure remains constant. In such cases, several different methods are available for modelling the coupling of the
system to its environment. For example, a range of different thermostats may be used.

For systems at thermal equilibrium, the results of molecular dynamics simulations depend on the choice of thermostat. However,
there is an established knowledge base as to which aspects of the systems are independent of this choice (for example, free
energies), and under what circumstances other aspects should be affected only mildly (for example, dynamical correlation functions
depend weakly on the choice of thermostat if the coupling to the environment is weak~\cite{allenTildesley} (Section~7.4.1)). This~knowledge is based on theoretical insights---for example, one typically uses thermostats 
that do not affect the invariant measure
(Boltzmann distribution) of the system and (if possible) also preserve the microscopic time-reversal symmetry of the equilibrium
state.

For systems that are far from thermal equilibrium, the situation is more complicated. Such systems are important and are
increasingly being modelled by molecular dynamics: see for example~\cite{hagan16emerge,nikolai16,martens16,Bouzid2017}. To~study
general features of such states, one might consider non-equilibrium steady states in which a material is simultaneously coupled to
two heat baths with different temperatures; or systems that are relaxing slowly towards an equilibrium; or systems in which some
variables are conditioned to take a non-typical value. In all these cases, the choice of thermostat (or barostat) can
significantly affect the dynamical behaviour, and it is not clear what choice is appropriate when modelling any specific system.
In particular, the invariant measure is (in general) no longer a Boltzmann distribution, and~time-reversal symmetry is broken, so
there are fewer principles available to constrain the design of suitable molecular dynamics models.

The same issues arise---even more noticeably---when proposing highly-simplified models of molecular dynamics systems. For
example, in Markov State Models (MSMs) of biomolecules~\cite{chodera07,chodera14}, one~represents a large molecule by a number of
discrete states, with Markovian transitions between them. In equilibrium, the relevant transition rates are constrained by the
principle of detailed balance (at least as long as the states depend only on a systems' configuration and not on its momenta).
Out of equilibrium, there are fewer general rules, although the modern theory of stochastic thermodynamics~\cite{Seifert2012rmp}
does address how physically-observable quantities like heat and work can be related to transition rates in simplified
(coarse-grained) stochastic models. As non-equilibrium systems are studied increasingly widely, we argue that general principles
for the design and interpretation of model systems is becoming increasingly important.

\subsection{Outline}

In this article, we analyse dynamical fluctuations in stochastic models of molecular systems. The~stochastic element of these
models represents the coupling of our system to its environment, and~the states in the models represent the coordinates and
momenta of the molecular system. We~review and extend recent work which showed how general symmetries and geometrical properties
govern dynamical fluctuations in these models, concentrating particularly on rare events (large deviations from the typical
behaviour). We propose that these general principles can be useful when building models of non-equilibrium states, since they
constrain the range of possible behaviour for different kinds of systems.

Our results are based on two recent developments, both of which focus on the key role of dissipation and the breaking of
time-reversal symmetry (a key concept in stochastic thermodynamics).

We first consider rare events in which an equilibrium (time-reversal symmetric) system spontaneously maintains a large current,
over a long period. It was argued in~\cite{JackEvans} that these events are free from dissipation, in contrast to typical
non-equilibrium steady states. In Section~\ref{sec:pt}, we review this argument and present some new examples that illustrate the
operation of the general principle. In particular, we focus on the role and definition of dissipation and entropy production in
these rare events. (Our results also have implications for the Maximum Caliber hypothesis~\cite{Dill2013} for building models of
non-equilibrium systems.)

The second part of this paper concerns fluctuations in irreversible Markov processes, and their analysis in terms of forces and
currents in the space of probability distributions. These currents and forces can be decomposed into reversible
(equilibrium-like) and ``non-equilibrium'' (irreversible) parts. Moreover, these two contributions to the force obey a kind of
orthogonality relation, which~has consequences for the non-equilibrium fluctuations. In Section~\ref{sec:orth}, we review recent
results in this direction, and we present a new application to systems described by a Hamiltonian evolution, coupled to a
thermostat. In contrast to the diffusive (overdamped) systems discussed previously, we argue that the different terms in the
theory have slightly different physical interpretations. We discuss the role of dissipation in that case.

\section{Definitions and Preliminaries}
\label{sec:prelims}

In this section, we collect several theoretical results needed in the following. They are primarily based on the theory of
stochastic thermodynamics, as reviewed in~\cite{Seifert2012rmp}.

\subsection{Model: Conservative Forces}

We consider a Hamiltonian system coupled to a thermostat at temperature $T$. There are $N$ particles moving in $d$ dimensions: we
denote their co-ordinates by $q=(q^i)_{i=1}^n$, with $n=Nd$. Each co-ordinate takes values on a circle of perimeter $L$: we take
$q^i\in\Lambda$ with $\Lambda :=[-L/2,L/2]$; the points $q^i=\pm L/2$ are identified with each other. The conjugate momenta are
$p=(p^i)_{i=1}^n$, so that $p\in\mathbb{R}^n$. \linebreak Define $\Omega := \Lambda^n \times \mathbb{R}^{n}$ as the phase space. All
particles have the same mass $m=1$ (cases where not all masses are equal can be analysed similarly, but we concentrate here on the
simplest case). The~Hamiltonian is
\begin{equation}
 H(q,p) = V(q) + \frac12 \sum_i (p^i)^2 , 
 \label{e2}
\end{equation}
where $V$ is the potential energy that depends only on the co-ordinates $q$. 

The system is coupled to a heat bath at temperature $T$ (we set Boltzmann's constant $k_{\rm B}=1$). We~assume that the coupling
of particle $i$ is independent of the co-ordinates, so the (stochastic) equations of motion are
\begin{equation}
 \dd q^i_t = p^i_t \dd t,
 \qquad
 \dd p^i_t = -\frac{\partial V}{\partial q^i} \dd t + \dd b^i_t . 
 \label{equ:ham-eom}
\end{equation}
The coupling of particle $i$ to the heat bath appears through the stochastic force
\begin{equation}
 \dd b^i_t = - \gamma p^i \dd t + \sqrt{2\gamma T} \dd W_t^i ,
 \label{equ:dft}
\end{equation}
where $\gamma$ is a friction constant and $\mathrm{d}W_t^i$ is a standard Brownian noise. (The Brownian noises $\mathrm{d}W_t^i$,
$\mathrm{d}W_t^j$ etc.~are all independent.) The generalisation to the case where the friction depends on the co-ordinates is
straightforward but requires some heavier notation. These stochastic differential equations are equivalent to Langevin equations
in physics: see~\cite{JackEvans} (Equations (2) and (3)), and replace $(\dd W/\dd t)$ by $\eta$.
With this choice the invariant measure (steady state probability distribution) $\pi$ for the phase space point $(q,p)$
satisfies~\cite{vanK}
\begin{equation}
 \dd\pi(q,p) = \frac{1}{Z(T)}\exp[-H(q,p)/T]\, \dd(q,p), \qquad Z(T) = \int_\Omega \exp[-H(q,p)/T]\, \dd(q,p),
 \label{equ:gibbs}
\end{equation}
where $\dd(q,p)=\dd q \dd p$, so the integral runs over all of phase space. The notation $\dd \pi(q,p)$ indicates the
(infinitesimal) probability that the system is at the phase space point $(q,p)$; the associated probability density is
$\ee^{-H/T}/Z$.

\subsection{Energy Flow into the Heat Bath}

It is useful to also consider the flow of energy from the system to the heat bath. The energy $E$ in the heat bath obeys the
equation of motion
\begin{equation}
 \dd E_t = \sum_i -p^i \circ \dd b^i_t , 
 \label{equ:dE-ham}
\end{equation}
where the circle indicates a Stratonovich product. This product of a force and a velocity is the rate at which the particles do
work on their surrounding environment, which therefore corresponds with the heat transfer. Combining with \eqref{equ:ham-eom} one
has
\begin{equation}
\dd E_t = -\sum_i \left( \frac{\partial V}{\partial q^i} \circ \dd q^i_t + p^i_t \circ \dd p^i_t \right) = -\dd H_t . 
\label{equ:dE-dH}
\end{equation}
Hence, $\dd (H+E)=0$: the total energy $H+E$ is (strictly) conserved.

Note that the internal co-ordinates $(q,p)$ evolve independently of $E$: this energy is useful as a book-keeping tool, but it does
not affect the system's dynamics. Hence, the heat flow into the bath over the time interval $[t',t]$ can be recovered as
\begin{equation}
 Q(t',t) := \int_{t'}^t \mathrm{d}E_s.
\label{equ:def-Q}
\end{equation}
(Absolute values of the bath energy $E$ are not well-defined within this theory, but the heat transfer may be computed for any
trajectory.) We also have $\dd(H+E)=0$ and thus $Q(t',t) = H(q_{t'},p_{t'}) - H(q_t,p_t)$.

\subsection{Model: Non-Conservative Forces}

To include non-equilibrium steady states in our general setting, we replace the gradient force $-(\partial V/\partial q^i)$
in \eqref{equ:ham-eom} by a general force $f(q)$ that depends only on the co-ordinates $q$, but is not necessarily the gradient of
a potential. That is, we consider
\begin{equation}
 \dd q^i_t = p^i_t \dd t, 
 \qquad
 \dd p^i_t = f^i(q_t) \dd t + \dd b^i_t . 
 \label{equ:noneq-eom}
\end{equation}
The coupling to the heat bath is still given by \eqref{equ:dft} and the energy of the heat bath obeys \eqref{equ:dE-ham}; the~heat
flow is still given by \eqref{equ:def-Q}. However,~\eqref{equ:dE-dH} becomes
\begin{equation}
 \dd E_t = \sum_i \left( f^i(q_t) \circ \dd q^i_t - p^i_t \circ \dd p^i_t \right) .
 \label{equ:dE-not-dH}
\end{equation}
The invariant measure of this system is not known in general---we denote the steady state distribution by $\pi$ but there is no
analogue of \eqref{equ:gibbs}. In addition, in the steady states of conservative systems, one expects the average of $E$ to be
independent of time: $\mathbb{E}[Q(t',t)]=0$ in steady state. For the non-conservative forces considered here, one expects
$\mathbb{E}[Q(t',t)]>0$ for $t>t'$ (unless $f$ is the gradient of a potential): see also Section~\ref{sec:heat-time} below.

\subsection{Path Measures}

We consider trajectories of these models, over a fixed time interval $[0,\tau]$. Define $X_t=(q_t,p_t)$ as the state of the
system at time $t$ and let $X=(q_t,p_t)_{t\in[0,\tau]}$ be a sample path (trajectory). Note that the equation of motion for the
co-ordinates $q$ has no stochastic part, so all possible trajectories of this system have $\partial_t{q}^i=p^i$. In general, we
use $\Pc$ to indicate a path measure for such a system, with initial conditions sampled from the invariant measure $\pi$. In addition,
let $\Pc_{X_0}$ be the path measure with fixed initial condition $X_0$. Hence
\begin{equation}
 \dd\Pc(X)=\dd\Pc_{X_0}(X) \cdot \dd\pi(X_0) .
 \label{equ:P-init-cond}
\end{equation}

To obtain an explicit representation of the path measure, we rewrite \eqref{equ:noneq-eom} as
\begin{equation}
 \dd q^i_t = p^i_t \dd t,
 \qquad
 \dd p^i_t = {\gamma} w^i(q_t,p_t) \dd t + \sqrt{2\gamma T} \dd W_t^i 
\end{equation}
and denote the invariant measure of this system by $\pi$. In this case, the (infinitesimal) probability of trajectory $X$ is given (in the Stratonovich convention) by
\begin{equation}
 \dd \Pc_{X_0}(X) = \exp\left(-\frac{1}{4 T}\int_0^\tau \big[-2w_t\circ \dd p_t + {\gamma}w_t^2 \dd t 
  + {2\gamma T}(\nabla_p\cdot w_t) \dd t\big]\right) \dd \Pc^{\rm ref}_{X_0}(X) , 
 \label{equ:dPX-qp}
\end{equation}
where $w_t$ indicates $w(q_t,p_t)$, and $\Pc^{\rm ref}_{X_0}$ is a reference measure (corresponding to $p_t$ being a random walk with diffusion constant $\gamma T$, and $\dd q=p \dd t$ as an equality).
Expectation values with respect to such path measures can be obtained as $\mathbb{E}_{X_0}[G] = \int G(X) \dd P_{X_0}(X)$. In physics one might equivalently write a path integral ({again} in Stratonovich convention)~\cite{Seifert2012rmp,lecomte17path}
\begin{equation}
 \mathbb{E}_{X_0}[G] = \int G[X] \exp\left( -\frac{1}{4T}\int_0^\tau 
  \left[ \frac{1}{\gamma}(\partial_t{p}_t - {\gamma} w_t)^2 + {2\gamma T}(\nabla_p\cdot w_t) \right] \dd t\right) 
 \delta[ \partial_t q- p ] \, \mathcal{D}X(t) , 
\end{equation}
which has exactly the same meaning (with $\delta[ \partial_t q- p ]$ encapsulating the constraint that $\partial_t{q}_t=p_t$ for
all times $t$).

\subsection{Time-Reversal Symmetry and Relation to Heat Flow}
\label{sec:heat-time}

Let $\mathbb{T}$ be a time-reversal operator acting on paths, which reverses the arrow of time and the direction of all momenta.
That is, $(\mathbb{T}X)_t := (q_{\tau-t},-p_{\tau-t})$. For a single phase space point $(q,p)$ let $\mathbb{T}(q,p) :=(q,-p)$.
The time-reversibility of Hamiltonian evolution combined with the appropriate combination of forces in \eqref{equ:dft} means that
for conservative systems described by \eqref{equ:ham-eom}, the steady state has a time-reversal symmetry
\begin{equation}
 \dd \Pc(X) = \dd \Pc(\TT X).
 \label{equ:TRS}
\end{equation}
Moreover, using \eqref{equ:dPX-qp}, as it applies to systems described by \eqref{equ:ham-eom} or~\eqref{equ:noneq-eom}, it may be
verified that
\begin{equation} 
 Q(0,\tau) = T \log \frac{\dd\Pc_{X_0}(X)}{\dd\Pc_{(\TT X)_0}(\mathbb{T}X)} , 
 \label{equ:Q-PX-PTX}
\end{equation}
which relates the heat flow into the bath to the breaking of time-reversal symmetry, for these systems.
For conservative systems, combining ~\eqref{equ:gibbs},~\eqref{equ:P-init-cond} and~\eqref{equ:TRS} recovers
$Q(0,\tau)=H(X_0)-H(X_\tau)$, as~required since $\dd Q_t=-\dd H_t$ in that case. In general (for both conservative and
non-conservative systems), averaging \eqref{equ:Q-PX-PTX} with respect to $\dd\Pc(X)$ {---} which corresponds to initial conditions
taken from the invariant measure---one sees also that $\mathbb{E}_{\Pc}[Q(0,\tau)]\geq 0$: the average steady-state energy flow
into the bath is non-negative, and vanishes only if the system is time-reversal symmetric. For the connection to fluctuation
theorems, see~\cite{Seifert2012rmp}.

\section{Absence of Dissipation in Conditioned Ensembles of Trajectories}
\label{sec:pt}

This section builds on recent work by Jack and Evans~\cite{JackEvans}, concerning dissipation in certain trajectory ensembles.
{The motivation for that work was a hypothesis~\cite{evans04} that properties of non-equilibrium steady states can be
 inferred by analysing a particular class of rare fluctuations that occur at equilibrium. These are rare events in which time-averaged
 currents have non-typical values. This hypothesis can be motivated as a far-from-equilibrium generalisation of linear-response
 theory, with its associated fluctuation-dissipation theorems and Onsager reciprocity relations. Following~\cite{JackEvans}, we
 show that this hypothesis fails qualitatively, in that it does not correctly account for dissipation in the non-equilibrium
 steady states.}

\subsection{Parity Symmetry}

The results of this section apply to conservative systems whose Hamiltonian has a parity symmetry. The idea is that on inverting
some subset of the coordinates $\mathcal{S} \subseteq \{1,2,\dots,n\}$ and their conjugate momenta, the Hamiltonian is unchanged.
Hence, define a parity operator $\mathbb{P}$ that acts on paths as $(\mathbb{P}X)=(\tilde{q}_t,\tilde{p}_t)_{t\in[0,\tau]}$ where
$(\tilde{q}^i_t,\tilde{p}^i_t)=(-q^i_t,-p^i_t)$ for $i\in \mathcal{S}$ and $(\tilde{q}^i_t,\tilde{p}^i_t)=(q^i_t,p^i_t)$
otherwise. \linebreak In addition, $\mathbb P$ acts on single phase space points as $\mathbb{P}(q,p)=(\tilde{q},\tilde{p})$. We restrict in the
following to parity-symmetric Hamiltonians $H$ for which
\begin{equation}
 H(\tilde{q},\tilde{p}) = H(q,p) .
\end{equation}
In this case, models with conservative forces as in \eqref{equ:ham-eom} satisfy
\begin{equation}
 \dd\Pc(X) = \dd\Pc(\PP X) = \dd\Pc(\TT X) = \dd\Pc(\PP\TT X) .
 \label{equ:eq-PT}
\end{equation}

\subsection{Examples}
\label{sec:eg}
 
To illustrate our general arguments, we focus on a very simple example: consider a single particle moving on a circle. The system
is described by \eqref{equ:ham-eom} with a single co-ordinate $q$ {and} conjugate momentum $p$. We take $L=1$ and choose
$V(q)=-V_0 \cos (2\pi q)$ for some constant $V_0\geq 0$. Hence
\begin{equation}
 \dd q_t = p_t \dd t, \qquad \dd p_t = [-V'(q_t) - \gamma p_t ] \dd t + \sqrt{2\gamma T}\dd W_t .
 \label{equ:eg-eq}
\end{equation}

The parity symmetry in this case is simply $\PP(q,p)=(-q,-p)$. We also define a corresponding non-equilibrium system (of the form 
\eqref{equ:noneq-eom}) where a (constant) force $\fext$ drives the system around the circle. In this case
\begin{equation}
 \dd q_t = p_t \dd t, \qquad \dd p_t = [ \fext - V'(q_t) - \gamma p_t ] \dd t + \sqrt{2\gamma T}\dd W_t .
 \label{equ:eg-noneq}
\end{equation}
We analyse this example in Section~\ref{sec:PT-compare} below. { Figure~\ref{fig:eg} illustrates the system, and some of
 its~properties.}
 
\begin{figure}
\includegraphics[width=12cm]{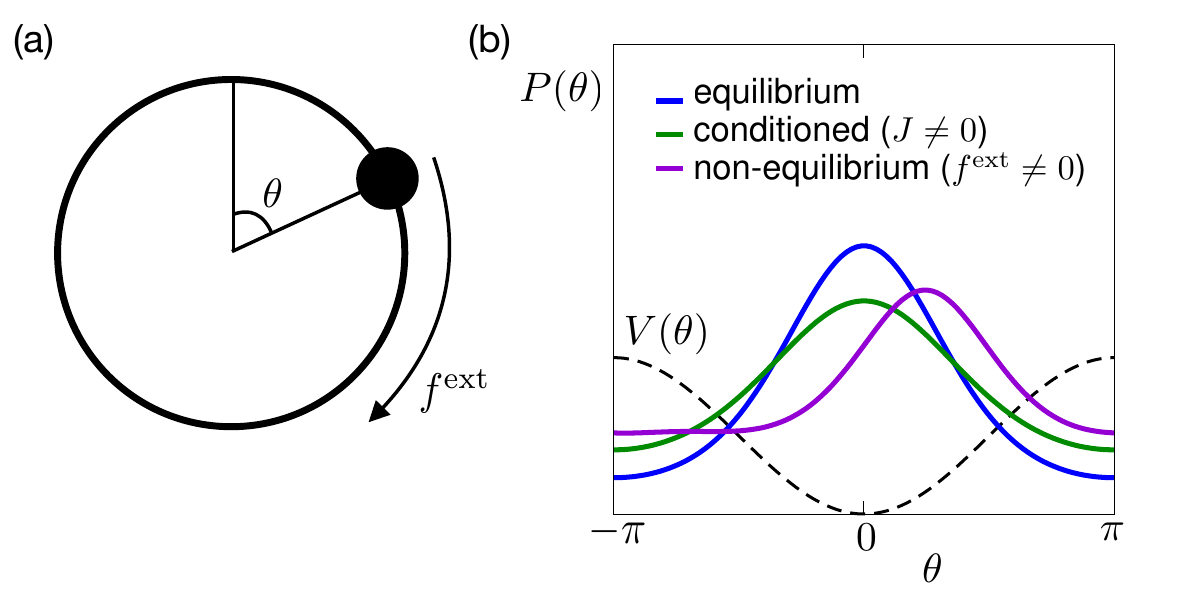}
\caption{(\textbf{a})~Illustration of the example system described in Section~\ref{sec:eg}. A particle moves on a circle, with co-ordinate
 $q=\theta/\pi$. There is a conservative force $-V'(q)$ and (possibly) an external force $f^{\rm ext}$ that drives the system
 around the circle. (\textbf{b})~Sketches showing the expected behaviour of the (marginal) probability density function of the angle
 $\theta$ in different states (see Section~\ref{sec:PT-compare}, also~\cite{JackEvans} (Figure~4)). The form of the potential is shown
 as a dashed line. At equilibrium ($f^{\rm ext}=0$) one has $P(\theta)\propto\exp(-V(\theta)/T)$. For $f^{\rm ext}>0$, the
 distribution favours positive $\theta$, so that the average of the gradient force $V'$ can balance the external force
 $f^{\rm ext}$, in the steady state. For conditioned ensembles with positive current, Symmetry~\eqref{equ:PTsym} requires
 that $P(\theta)$ remains symmetric under $\theta\to-\theta$, see Equation~\eqref{equ:PQ-sym}. However, to support the finite current
 $J$, the distribution of the momentum $p=\dot{q}=\dot\theta/\pi$ must have a distribution that favours $p>0$ (not shown).%
}
\label{fig:eg}
\end{figure}

For an application of these ideas in a more complicated example---fluid motion under shear---see~\cite{JackEvans}. There, the
particles move in two dimensions: they are confined in the vertical direction, but there are periodic boundaries in the horizontal
direction. Shear flow occurs when the upper and lower boundaries move (horizontally) with relative velocity $v$. In that case,
the parity $\mathbb{P}$ corresponds to a plane reflection, which reverses the direction of the shear flow, but leaves the
orthogonal directions unchanged (see Figure 1 in~\cite{JackEvans}).

\subsection{Currents and Fluxes}
\label{sec:fluxes}

Define $u(q,p):=\sum_i a_i(q) p^i$ as a general momentum, that changes its sign under time-reversal (here the $a_i$ are a set of
weight functions). We consider the time-integral of one such momentum, which~we identify as an (average) flux
\begin{equation}
 J^\tau(X) := \frac{1}{\tau} \int_0^\tau \sum_{i\in \cal S} a_i(q_t) p^i_t\, \mathrm{d}t.
 \label{equ:def-Jtau}
\end{equation}
We further assume that $a_i(\tilde q)=a_i(q)$: with this choice $J^\tau$ changes its sign under parity-reversal. Hence
\begin{equation}
 J^\tau(\mathbb{T}X)=-J^\tau(X)=J^\tau(\mathbb{P}X).
\end{equation}
As a shorthand for such an equation, we say that $J^\tau$ is ``odd'' in both $\TT$ and $\PP$. However, $J$ is even in the
combined symmetry operation $\PP\TT$, that is,
\begin{equation} 
 J^\tau(\PP\TT X)=J^\tau(X). 
 \label{equ:J-PT} 
\end{equation}

In general, we identify time-integrated quantities that are odd in $\TT$ as ``fluxes''. An important example is
$Q(0,\tau)=\int_0^\tau \mathrm{d}E_s$ (recall \eqref{equ:def-Q}), which is odd in $\TT$ but even in $\PP$. Hence $Q$ is odd under
$\PP \TT$: we write $Q(X)$ for the heat transfer associated to path $X$ so that
\begin{equation} 
 Q(\PP\TT X)=-Q(X). 
 \label{equ:Q-PT} 
\end{equation}
We refer to fluxes that are odd in $\PP$ (and hence even in $\PP\TT$) as ``transport fluxes'': we imagine that some quantity is
being transported through the system in a particular direction that is odd under parity. On the other hand, we refer to fluxes
that are even in $\PP$ (and hence odd in $\PP\TT$) as ``dissipative fluxes'': they are independent of the direction of transport.

\subsection{Ensembles and PT-Symmetry}

We consider a set of rare sample paths for which $J^\tau$ has a non-typical value $J$. In later sections, we~consider the limit
of large $\tau$, but for the moment, $\tau$ can take any value. Define a path distribution $\mathcal{P}^J$ that is conditioned on
this value of $J^\tau$ as
\begin{equation}
 \dd\mathcal{P}^J(X) := \dd\mathcal{P}(X | J^\tau(X) = J).
\end{equation}
This definition means that $\dd\mathcal{P}^J(X)=0$ if $J^\tau(X)\neq J$. In addition, if two trajectories $X,X'$ satisfy
$J^\tau(X)=J^\tau(X')=J$, then
\begin{equation}
 \frac{\dd\Pc^J(X)}{\dd\Pc^J(X')}
 = \frac{\dd\Pc(X)}{\dd\Pc(X')}.
 \label{equ:Pcond-ratio}
\end{equation}

Now, fix some $J\neq0$ and consider a trajectory $X$ with $J^\tau(X)=J$. Then, trajectories $\mathbb{P}X$ and $\mathbb{T}X$ both
have $J^\tau=-J$, so $\dd\mathcal{P}^J(\mathbb{P}X) = 0 = \dd\mathcal{P}^J(\mathbb{T}X) $. On the other hand, \eqref{equ:J-PT}
means that trajectory $\mathbb{PT}X$ has $J^\tau(\mathbb{PT}X)=J$ and \eqref{equ:eq-PT} implies that
$\dd\mathcal{P}(\mathbb{PT}X) = \dd\mathcal{P}(X)$. Hence, using \eqref{equ:Pcond-ratio}, one has
\begin{equation} 
 \mathcal{P}^J(\mathbb{PT}X) = \mathcal{P}^J(X) . 
 \label{equ:PTsym} 
\end{equation}
This symmetry of conditioned path ensembles mirrors the main result of~\cite{JackEvans} (which applies to a related set of
``biased'' path ensembles). (We {derived this symmetry for systems} described by \eqref{equ:ham-eom} but the discussion of this
section generalises immediately to any system as long as the path measure satisfies
$\mathcal{P}(\mathbb{T}X)=\mathcal{P}(X)=\mathcal{P}(\mathbb{P}X)$ and the current $J^\tau$ satisfies
$J^\tau(\mathbb{T}X)=-J^\tau(X)=J^\tau(\mathbb{P}X)$.)

\subsection{Observable Consequences}

Recall that the heat flow $Q(0,\tau)$ is odd under $\PP\TT$. Hence, the average value of $Q$ for the conditioned ensemble is
\begin{align}
 \mathbb{E}^J[ Q ] 
 & = \int Q(X) \mathrm{d}\mathcal P^J(X) 
  \nonumber \\
 & = \frac12 \int Q(X) \mathrm{d}\mathcal P^J(X) + \frac12 \int Q(\mathbb{PT}X) \mathrm{d}\mathcal P^J(\mathbb{PT}X)
  \nonumber \\
 & = \frac12 \int Q(X) \mathrm{d}\mathcal P^J(X) - \frac12 \int Q(X) \mathrm{d}\mathcal P^J(X)
  = 0 , 
  \label{equ:EQ-zero}
\end{align}
where the second equality is obtained by a change of integration variable and the third uses \eqref{equ:Q-PT} and~\eqref{equ:PTsym}. Hence, on average, no energy flows into the heat bath in the steady state of the conditioned ensemble, even
though a finite current $J$ is flowing.

In the nomenclature of Section~\ref{sec:fluxes}, this argument can be used to show that all dissipative fluxes vanish in the
conditioned ensemble. Hence we say that \emph{{the conditioned ensemble is free from dissipation}}. On the other hand, note that
all co-ordinates $q^i$ are even under $\mathbb T$. In this case, the derivation \eqref{equ:EQ-zero} may be used to show that the
averages of all co-ordinates that are odd in $\mathbb{P}$ must vanish in the conditioned ensemble, see~\cite{JackEvans} for
specific examples.

\subsection{Large Deviation Principle and Auxiliary Process}
\label{sec:aux}

We now consider the limit of large time $\tau$, in which case the probability that a sample path has $J^\tau(X)=J$ is governed by
a large deviation principle at ``level-1'' (in the nomenclature of Donsker and Varadhan). We write this
as~\cite{chetrite2015condition}
\begin{equation} 
 \mathrm{Prob}[ J^\tau \approx J ] \asymp \exp[ -\tau I(J) ] ,
 \label{equ:j-ldp-lev1} 
\end{equation}
where $I$ is the rate function. One has $I(J)\geq0$; also, if $I(J)>0$, then the probability of current $J$ is suppressed
exponentially as $\tau\to\infty$. These events are clearly very rare. The conditioned distribution $\mathcal{P}^J$ is not easy to
analyse. To make further progress, {it is useful to define an ``auxiliary'' Markov
 process~\cite{maesEPL2008,JackSollich2010,popkov10,nemoto11,chetrite2015condition,JackSollich2015} whose steady state path
 measure is close to $\mathcal{P}^J$:} see~\cite{chetrite2015condition} for a comprehensive discussion. To do so, we first define
a scaled cumulant generating function
\begin{equation} 
 \psi(s) := \sup_J [ sJ - I(J) ] 
 \label{equ:def-psi-scgf} 
\end{equation}
and define $j(s) = \psi'(s)$, such that $j$ is a monotonically increasing function of $s$, with inverse $s^*=j^{-1}$. Assuming
that $I$ is convex (that is, there are no dynamical phase transitions~\cite{Touchette2009}), then
\begin{equation} 
 I(J) = \sup_s [ sJ - \psi(s) ] , 
\end{equation}
and the value that achieves the supremum is $s^*(J)$. Assuming that the original process of interest is given
by \eqref{equ:ham-eom}, define an $s$-dependent auxiliary process as
\begin{align}
 \dd q^i_t = p^i_t \dd t , \qquad
 \dd p^i_t = \dd \hat{b}^i_t -\frac{\partial V}{\partial q^i} \dd t - \gamma T \frac{\partial G^s}{\partial p^i} \dd t 
 \label{equ:eom-G}
\end{align}
for some function $G^s\colon\Omega\to\mathbb{R}$ to be specified below. The force $\hat{b}^i_t$ has the same statistical
properties as $b^i_t$ in \eqref{equ:noneq-eom}, but we use a different notation because we are going to make a mapping between
sample paths of this auxiliary process and sample paths of the original process \eqref{equ:ham-eom}. At the level of sample paths,
$b\neq \hat{b}$. We identify $\gamma(\partial G^s/\partial p^i)$ as a ``control
force''~\cite{JackSollich2015,chetrite2015control} that realises the required flux $J$ 
{(see also~\cite{nemoto11}).}

Since the sample paths of the auxiliary process should be as close as possible to those of the original process, we use the fact
that the heat current in the original process is a deterministic function of $(p,q)$ and satisfies \eqref{equ:dE-dH}. Hence, that
equation also is used to compute the energy $E_t$ for the auxiliary~process.

The determination of a suitable function $G^s$ is described in~\cite{chetrite2015condition}. Briefly, let $\cal L$ be the
generator of the process \eqref{equ:ham-eom}. Then $\exp(-G^s)$ solves the eigenvalue equation
\begin{equation}
 \left[ {\cal L} + s\sum_{i\in \cal S} a_i(q) p^i \right] \exp(-G^s) = \psi(s) \exp(-G^s) , 
\end{equation}
where the coefficients $a_i$ are those appearing in \eqref{equ:def-Jtau} and the eigenvalue $\psi(s)$ coincides
with \eqref{equ:def-psi-scgf}. Under~these conditions, let $\mathcal P^{\rm aux}_s$ be the path measure of this $s$-dependent
process. Then
\begin{equation} 
 \mathcal P^{\rm aux}_{s^*(J)} \approx \mathcal P^J 
 \label{equ:pcond-paux} 
\end{equation}
in the sense that the relative entropy between these two distributions is $o(\tau)$~\cite{chetrite2015condition}.

For our purposes, this result has two important implications. First, the analysis of the (intractable) probability measure $P^J$
has been reduced to analysis of the auxiliary model, which is often easier. Second, it means that the physical behaviour
associated with the conditioned ensemble $P^J$ can be reproduced by the stochastic process \eqref{equ:eom-G}. In particular, this
auxiliary process achieves current flow without dissipation. This unusual situation can be achieved only with the aid of a
``control potential'' $G^s$, which in general has a complex dependence on $q$ and $p$: see Section~\ref{sec:PT-compare}. In addition,
comparing \eqref{equ:ham-eom} and~\eqref{equ:eom-G}, one sees that $\dd {b}^i_t$ corresponds to
$\dd \hat{b}^i_t - \gamma T(\partial G^s/\partial p^i) \dd t$. The interpretation of this fact is that when one considers the
conditioned process, the stochastic noises that appear in the definition of the original process do not have a mean value of zero
any more: in fact the $\dd W^i_t$ that appears in the definition of $\dd {b}^i_t$ has a mean value proportional to
${\partial G^s}/{\partial p^i}$ once the conditioning is applied. That is, one can think of the control force as a bias on the
noise that is induced by the conditioning.

\subsection{Example System: Comparisons between the Auxiliary Process and Other Physical Ensembles}
\label{sec:PT-compare}

In this section, we compare the auxiliary process defined above with two other physical processes, in order to explore in more
detail the nature of dissipation. We use the example system \eqref{equ:eg-eq} to illustrate the relevant ideas. We take
$J^\tau = \frac{1}{\tau} \int_0^\tau p_t \dd t$.

In the special case where there is no potential ($V_0=0$), the auxiliary process can be obtained exactly. The generator $\cal L$
acts on functions $g\colon\Omega\to\mathbb{R}$ as ${\cal L}g = (p\cdot\nabla_q - \gamma p\cdot\nabla_p + \gamma T\nabla_p^2) g$.
One~finds $G_s(p,q) = (-sp/\gamma)$ and $\psi(s)=(Ts^2/\gamma)$. In that case, the equations of motion \eqref{equ:eom-G} for the
auxiliary process coincide with the non-equilibrium system~\eqref{equ:eg-noneq}, with $\fext = sT$. However, as noted above, the
heat flow in the auxiliary process is given by~\eqref{equ:dE-dH}. On the other hand, the heat flow in the non-equilibrium system
is given by~\eqref{equ:dE-not-dH}, with $f=\fext$, a constant. In this case, it is easily verified that for long trajectories
($\tau\to\infty$), the auxiliary process (and hence the conditioned process) have $\frac{1}{\tau}\mathbb{E}[Q(0,\tau)]=0$ while
the non-equilibrium process has $\frac{1}{\tau}\mathbb{E}[Q(0,\tau)]=(\fext)^2/\gamma$.

We emphasise that this case $V_0=0$ is a special one {---} if one inspects the statistics of the particle trajectories (that is,
$(q_t,p_t)_{t\in[0,\tau]}$) then it is not possible to determine whether one is observing the non-equilibrium
system~\eqref{equ:eg-noneq} or the conditioned equilibrium system based on~\eqref{equ:eg-eq}. On the other hand, if~one observes
(by some physical measurement) the heat flow into the reservoir for these two cases, then one sees that the conditioned process
has no dissipation (no net heat flow), but the non-equilibrium process does have a finite rate of heat flow into the environment.

In the general case $V_0>0$, the two ensembles are more easily distinguished. For example, one~may show {(see~\eqref{equ:PQ-sym} below)} that
$\mathbb{E}^{J}(q)=0$ but the steady state of the non-equilibrium process has $\mathbb{E}(q)>0$ if $\fext>0$. In that case, the
physical situation is that the external force $\fext$ drives the system away from the potential minimum (at $q=0$); once the
particle has reached the maximum then it wraps around the circle and falls back to the minimum, and the work that was done by the
external force is dissipated as heat in the bath. On the other hand, the physical interpretation of the conditioned process is
that the particle borrows energy from the heat bath in order to overcome the barrier, before returning that energy to the heat
bath as it falls back down again. 
{For explicit computations on a similar system in the overdamped limit, see~\cite{touchette16circle}.}

The physical interpretation of the auxiliary process in this case is that the control force $-\gamma T(\partial G/\partial p)$
does work to push the system away from the minimum, but this work is not dissipated as heat in the bath: instead the control force
acts to slow down the particle as it falls back to the minimum, in such a way as to avoid any dissipation. We expect that this
requires complex velocity-dependent forces that are not expected in typical equilibrium systems. One may imagine that the control
potential is applied by a kind of Maxwellian demon, that has full control over all aspects of the particle motion, and hence can
avoid the usual expectations of thermodynamics, that persistent particle currents should be accompanied by dissipation.

Based on the numerical results of~\cite{JackEvans} and the symmetries of the problem, we illustrate in Figure~\ref{fig:eg}b 
how the parity-time (PT) symmetry affects the conditioned steady state of the example problem discussed in Section~\ref{sec:eg}. One observes a 
qualitative difference between the conditioned steady state and the non-equilibrium steady state that is observed when $f^{\rm ext}>0$. To see this, note that
if a co-ordinate $q^i$ is odd under the parity operation $\PP$ then its marginal distribution (probability density) $P^J_i$ is necessarily symmetric, in the conditioned steady state. This steady-state distribution is evaluated at some appropriate time $t$: for example, consider the limit $\tau\to\infty$ with $t=\alpha \tau$ for some $\alpha\in(0,1)$. Then
\begin{align}
 P_i^J(q) 
 & = \int \delta(q-q^i(X_t)) \mathrm{d}\mathcal P^J(X) 
  \nonumber \\
 & = \int \delta(q-q^i(\PP X_{\tau-t})) \mathrm{d}\mathcal P^J(\mathbb{PT}X)
  \nonumber \\
 & = \int \delta(q+q^i(X_{\tau-t})) \mathrm{d}\mathcal P^J(X)
  = P_i^J(-q) , 
  \label{equ:PQ-sym}
\end{align}
where $q^i(X_t)$ is the value of co-ordinate $q^i$ at time $t$ in trajectory $X$, the second line is a change of integration variable $X\to\PP\TT X$, the third uses~\eqref{equ:PTsym} and that $q^i$ is odd under $\PP$. The last equality uses that $P^J_i(q)$ is independent of the parameter $\alpha$. Hence one has also $\mathbb{E}^J(q^i)=\int_{-L/2}^{L/2} P_i^J(q) \mathrm{d}q=0$.

Returning to the case $V_0=0$, there is one other instructive comparison. We imagine that the noise force $\dd b$
in~\eqref{equ:eg-eq} comes from friction between the particle and a surrounding solvent, but now imagine that the solvent is
moving with constant velocity $v$. We refer to this as a system with advection\linebreak (of the particle, by the solvent). In this case
the equations of motion are obtained by applying a Galilean transformation to~\eqref{equ:eg-eq}, which yields
\begin{equation}
 \dd q_t = p_t \dd t, \qquad \dd p_t = \dd b_t^v , \qquad \dd E_t = (p_t-v) \circ \dd b_t^v , 
 \label{equ:eom-advect}
\end{equation}
where in this case
\begin{equation}
 \dd b_t^v = \gamma ( v - p_t ) \dd t + \sqrt{2\gamma T} \dd W_t.
\end{equation}

In this equation, the first term on the right hand side comes from friction with the moving solvent. In this case one may verify
$\dd E_t = -dH_t$ with $H=(p_t-v)^2/2$. The steady state has $(\dd/\dd t)\mathbb{E}(H_t)=0$ and so there is no heat flow into the
bath: $\frac{1}{\tau}\mathbb{E}[Q(0,\tau)]=0$.

\subsection{Formulae for Heat Flow in Terms of Path Probabilities}

Recall~\eqref{equ:Q-PX-PTX}, which connects the heat flow in a trajectory with the ratio of probabilities of forward and backward
paths, for the systems described by~\eqref{equ:ham-eom}--\eqref{equ:noneq-eom}. It follows from that equation that
$\frac{1}{\tau}\mathbb{E}[Q(0,\tau)]>0$ if, for typical trajectories $X$, $\dd\Pc_{X_0}(X)$ differs significantly from
$\dd\Pc_{(\TT X)_0}(\TT X)$. However, the results of the previous section show clearly that systems with advection and
conditioned ensembles (and auxiliary process) violate~\eqref{equ:Q-PX-PTX}, in that there is breaking of time-reversal symmetry,
but~no heat flow.

For the case with advection, the solution to this apparent paradox~\cite{Speck2009} is that one should
replace~\eqref{equ:Q-PX-PTX} by the alternative formula
\begin{equation}
 Q(0,\tau) = T \log \frac{\dd\Pc^v_{X_0}(X)}{\dd\Pc^{-v}_{(\TT X)_0}(\mathbb{T}X)} , 
\end{equation}
where $\Pc^v$ is the path probability distribution for the system with solvent velocity $v$, and similarly $\Pc^{-v}$ has solvent
velocity $-v$. It may be checked directly from the path probabilities~\eqref{equ:dPX-qp} that this gives the correct heat
transfer in our case. Our inference from~\cite{Speck2009} is that one should not regard~\eqref{equ:Q-PX-PTX} as a fundamental
formula for heat flow: one should instead compute the heat transfer to the bath directly using~\eqref{equ:eom-advect} and then
derive the corresponding formula in terms of path probabilities.

Based on that assumption, it is easily verified that for the conditioned ensembles as defined here, one should take
\begin{equation}
 Q(0,\tau) = T \log \frac{\dd\Pc^{J}_{X_0}(X)}{\dd\Pc^{-J}_{(\TT X)_0}(\mathbb{T}X)}
\end{equation}
or, equivalently, 
\begin{equation}
 Q(0,\tau) = T \log \frac{\dd\Pc^{J}_{X_0}(X)}{\dd\Pc^{J}_{(\PP\TT X)_0}(\PP\mathbb{T}X)},
\end{equation}
as proposed in~\cite{JackEvans}.

\subsection{Outlook}

We summarise the outcomes of this analysis as follows. First, conditioning on fluxes that are odd in $\PP$ leads to
dissipation-free ensembles, in the sense that no heat flows from the system into its environment. Second, the behaviour observed
in this ensembles can be reproduced by auxiliary models, but this requires an ``optimal control'' potential $G^s$ that (typically)
depends in a complicated way on all co-ordinates in the system, and does not correspond to a simple physical driving force. { The~fact that such forces tend to have a complex dependence on the system's state has been remarked before~\cite{JackSollich2010,nemoto11,jack14east,jack15hyper}. In the present context, our results help to rationalise this fact: these forces act to drive currents without inducing dissipation, so they must inevitably be very different from driving forces that appear in typical physical systems.}
Third,
entropy production (in the environment) can (in~these situations) be directly computed in terms of an energy flow, which helps to
clarify what is the appropriate formula for obtaining $Q$ in terms of path probabilities.

{To see the consequences of these results, we focus on the comparison between non-equilibrium steady states (e.g., in the example~\eqref{equ:eg-noneq}), and conditioned ensembles of trajectories. In both cases, currents flow through the system, but only the conditioned ensemble respects the $\PP\TT$ symmetry~\eqref{equ:PTsym}. This property of the conditioned ensemble has its origin in the symmetries of the underlying dynamics, which still have implications for rare fluctuations in which large currents are sustained over long time periods. In response theory for equilibrium states, it is familiar that the same symmetries place strong constraints on linear responses, leading (for example) to Onsager reciprocity and fluctuation-dissipation theorems~\cite{Seifert2012rmp}. However, the far-from-equilibrium steady states considered here do not retain any such symmetries---the connection between spontaneous fluctuations and responses to perturbations has broken down, as do the usual fluctuation-dissipation theorems. As discussed in~\cite{JackEvans}, this difference between responses and spontaneous fluctuations leads to the failure of maximum entropy (or maximum caliber) approaches such as that of~\cite{evans04}.}

\section{Orthogonality of Forces and Currents in non-Equilibrium Systems}
\label{sec:orth}

In this section, we discuss a different set of symmetry properties of dynamical fluctuations in systems with non-conservative
dynamics, coupled to a heat bath. The idea is to decompose forces in the system into two pieces, according to their behaviour
under time-reversal. This leads to a decomposition of the heat flow into two contributions---housekeeping heat and excess heat.
It~also leads to a decomposition of probability currents which has a geometrical interpretation: the current has two orthogonal
components, one of which can be attributed to a free energy gradient. For~overdamped systems, these results are familiar from the
theory of stochastic thermodynamics~\cite{Seifert2012rmp} and from the Macroscopic Fluctuation Theory~\cite{Bertini2015}. We will
show that for systems with momenta, the construction is slightly more complicated, and we discuss the resulting decompositions and
their geometrical~interpretations.

\subsection{Overdamped Diffusions}
\label{sec:over}

We first review the situation in overdamped systems described by stochastic differential equations ({SDEs})
or first-order Langevin equations. We summarise relevant
results from stochastic thermodynamics~\cite{Seifert2012rmp} and from Macroscopic Fluctuation Theory~\cite{Bertini2015}. The
physical significance of these results is summarised in Section~\ref{sec:over-phys}.

\subsubsection{Model}

We consider a system with state $x_t=(x^i_t)_{i=1}^n$, which takes values in a space $\Gamma\subseteq\mathbb{R}^n$. It evolves in
time as
\begin{equation}
 \dd x^i_t = v_i(x_t) \mathrm{d}t + \sqrt{2\gamma_i T} \dd W_t^i , 
 \label{equ:sde-over}
\end{equation}
where we introduced a set of noise intensities $(\gamma_i)_{i=1}^n$, one for each coordinate. Assuming that all the $\gamma_i$
are finite, we identify forces that drive the particle motion as
\begin{equation}
 f_i(x) = v_i(x) / \gamma_i .
 \label{equ:f-gamma-v}
\end{equation}

Comparing with~\eqref{equ:noneq-eom}, one sees that $\gamma$ plays the role of a noise intensity in both systems. One way to
arrive at~\eqref{equ:sde-over} is to consider the overdamped limit of~\eqref{equ:noneq-eom}; note however that on taking this
limit, the~noise intensity $\gamma_i$ in~\eqref{equ:sde-over} does \emph{{not}} correspond to the friction constant $\gamma$ 
in~\eqref{equ:noneq-eom}.

The heat transfer to the environment is~\cite{Seifert2012rmp} (Equation 16)
\begin{equation}
 \dd E_t = f(x_t) \circ \dd x_t.
 \label{equ:dE-diff}
\end{equation}

The path measure for this system is given by an analogue of~\eqref{equ:dPX-qp}, which is
\begin{equation}
 \dd \Pc_{X_0}(X)\propto \exp\left(-\frac{1}{4T}\int_0^\tau \big[-2f_t\circ \dd x_t + f_t \cdot \hat\gamma f_t \dd t 
  + 2T \nabla\cdot (\hat\gamma f_t) \dd t\big]\right) \dd \Pc^{\rm ref}_{X_0}(X) , 
 \label{equ:dPX-diff}
\end{equation}
where we write $f_t=f(x_t)$ for compactness of notation, $\hat\gamma$ is a diagonal matrix with elements $(\gamma_i)$, and~$\Pc^{\rm ref}$ corresponds to a random walk for $x_t$ with ``diffusion matrix'' $\hat\gamma T$. The SDE~\eqref{equ:sde-over} is
associated with a Fokker-Planck equation~\cite{vanK} that describes the evolution of a probability density $\rho$ on $\Gamma$, as
\begin{equation}
 \partial_t \rho = -\nabla\cdot J(\rho), \qquad J(\rho) = \hat\gamma (f\rho - T\nabla\rho) .
 \label{equ:fp-over}
\end{equation}

Finally, for a general current $j\colon\Gamma\to\mathbb{R}^n$ and a vector field $F\colon\Gamma\to\mathbb{R}^n$, it is useful to
define
\begin{equation}
 \langle j,F \rangle := \int_\Gamma (j \cdot F) \dd x .
 \label{equ:dual-Gamma}
\end{equation}

\subsubsection{Time Reversal and Heat Transfer}

Define a time-reversal operation $\TT_0$ which reverses time but does not change any coordinates or momenta, as is appropriate for
overdamped dynamics. That is, for paths $X$ on the time interval $[0,\tau]$, we take $(\TT_0X)_t=(X)_{\tau-t}$. Now define an
adjoint dynamics~\cite{Bertini2015} for which the path measure is $\Pc^*$, with
\begin{equation}
 \dd\Pc^*(X)=\dd\Pc\left(\TT_0 X\right) .
 \label{equ:P*}
\end{equation}
That is, the steady probability of a particular path under the adjoint dynamics is equal to the corresponding probability of the
time-reversed path, under the orginal dynamics. By considering paths with $\tau\to0$, one sees that the invariant measure
associated with the adjoint process is the same as that of the original process, $\pi^*=\pi$.

The equations of motion of the adjoint process may be derived, either directly from~\eqref{equ:dPX-diff} or using the
Fokker-Planck equation~\eqref{equ:fp-over}. This latter approach is outlined in Appendix~\ref{sec:app-over}. We summarise the
result: let the invariant measure of the process be $\pi$, with
\begin{equation} 
 \dd\pi(x) = \frac{\ee^{-U(x)}}{Z_0} \dd x, \label{equ:inv-over} 
\end{equation}
and $Z_0=\int_\Gamma \ee^{-U(x)}\dd x$ for normalisation. The ``potential'' $U$ can be obtained by solving a partial differential
equation: see~\eqref{equ:U-over}. Then the adjoint process has equation of motion~\eqref{equ:sde-over}, with $f_i$
of~\eqref{equ:f-gamma-v} replaced by
\begin{equation} 
 f^*_i=-\left(2T\frac{\partial U}{\partial x^i}+f_i\right) . 
 \label{equ:f*-over} 
\end{equation}
If $f_i=-(\partial V/\partial x^i)$ for some potential $V$ then~\eqref{equ:sde-over} corresponds to the overdamped limit of a
conservative system, the invariant measure is $\dd \pi(x) \propto \ee^{-V(x)/T} \dd x$, and $f=f^*$. Hence the original and
adjoint processes concide, and the system is time-reversal symmetric: $\dd\Pc(\TT_0X)=\dd\Pc^*(X)=\dd\Pc(X)$.

Defining $\dd\Pc_{X_0}^*(X):=\dd\Pc^*(X)/\dd\pi^*(X_0)$ by analogy with~\eqref{equ:P-init-cond}, and recalling that $\pi^*=\pi$,
the~analogue of~\eqref{equ:Q-PX-PTX} in this system is
\begin{equation}
 Q(0,\tau) = \int_0^\tau \dd E_t = T \log \frac{\dd\Pc_{X_0}(X)}{\dd\Pc_{X_\tau}(\mathbb{T}_0X)} ,
 \label{equ:Q-PX-diff}
\end{equation}
which may be verified directly from~\eqref{equ:dE-diff} and~\eqref{equ:dPX-diff}. Now define
\begin{equation}
 Q^{\rm hk}(0,\tau) = T \log \frac{\dd\Pc_{X_0}(X)}{\dd\Pc_{X_0}^*(X)},
 \label{equ:Q-hk-diff}
\end{equation}
which is known as the housekeeping heat~\cite{Seifert2012rmp}. 
Note that since $\pi=\pi^*$ one could equivalently define
$Q^{\rm hk}(0,\tau) = T \log [\dd\Pc(X)/\dd\Pc^*(X)]$, but we choose to use path probabilities conditioned on their initial
states, for later convenience. Using~\eqref{equ:P-init-cond},~\eqref{equ:inv-over} as well as $\pi^*=\pi$, one sees that
\begin{equation}
 Q(0,\tau) = Q^{\rm hk}(0,\tau) + T \log \frac{\dd\pi(X_\tau)}{\dd\pi(X_0)} = Q^{\rm hk}(0,\tau) - T\int_0^\tau \dd U_t.
 \label{equ:QQ-diff}
\end{equation}
That is, the total heat has two components: the final term on the right-hand-side is related to the difference in probability
between initial and final states and says that heat is transferred to the bath as the system relaxes towards more likely
configurations. The other contribution $Q^{\rm hk}(0,\tau)$ is the additional heat flow that is not associated with relaxation
towards more likely states. This is a dissipative heat flow and represents energy input from external forces that is not available
for doing work, but must be expended in order ``to do the housekeeping''. In steady states
$\mathbb{E}(Q)=\mathbb{E}(Q^{\rm hk})$: the~only contribution to the (average) heat flow is the housekeeping heat.

\subsubsection{Splitting of the Force According to Time-Reversal}

Define 
\begin{equation}
 f^S := \frac12 (f + f^*) = -T\frac{\partial U}{\partial x^i}, \qquad f^A:=\frac12 (f-f^*) = f+T\frac{\partial U}{\partial x^i}.
 \label{equ:fs-fa-over}
\end{equation}
Since the adjoint process corresponds to a time-reversed dynamics, one sees that the force $f^S$ is even (symmetric) under
time-reversal, while $f^A$ is odd (anti-symmetric). From~\eqref{equ:dE-diff} and~\eqref{equ:QQ-diff}, one then sees that
\begin{equation}
 Q^{\rm hk}(0,\tau) = \int f^A(x_t) \circ \dd x_t .
 \label{equ:Qhk-fA-diff}
\end{equation}
That is, the housekeeping heat is associated with the anti-symmetric force, while the remaining (excess) heat is associated
with the symmetric force.

We note that for consistency of~\eqref{equ:Qhk-fA-diff} with~\eqref{equ:dPX-diff} and~\eqref{equ:Q-hk-diff} one must also have
\begin{equation}
 f^A \cdot \hat\gamma f^S + T\nabla \cdot (\hat\gamma f^A) = 0.
 \label{equ:fs-fa-orth}
\end{equation}
This may be verified using~\eqref{equ:fs-fa-over} together with~\eqref{equ:U-over}. It is also equivalent to
$\ddiv(f^A \ee^{-U})=0$, which means that if $\rho\propto\ee^{-U}$ is the invariant density, then the corresponding probability
current $\rho f^A$ is divergence free, and therefore does not transport any density: see for example~\cite{Spiliopoulos2016}. It
follows that for systems of the form~\eqref{equ:sde-over} and \eqref{equ:f-gamma-v}, one may replace the force $f$ by
$f^\lambda=f^S+\lambda f^A$ and the invariant measure is independent of $\lambda$.

\subsubsection{Large Deviation Principle for Many Copies of the System}
\label{sec:ldp-over}

So far we have considered dynamical fluctuations at the level of individual sample paths. To~gain further insight, it is useful
to consider a large deviation principle (LDP) that appears when we consider $M$ independent copies of our system, with a limit
$M\to\infty$. The resulting LDP is of the same form as those considered in Macroscopic Fluctuation Theory (MFT). This allows us
to identify an orthogonality relation between two contributions to the probability current $J$ that appears
in~\eqref{equ:fp-over}: these~two contributions originate from the splitting $f=f^S+f^A$.

To this end, define the empirical density $\rho^M$ such that $\int_V \rho^M \dd x$ is the number of copies of the system whose
positions $x$ are inside any volume $V\subset\Gamma$. Similarly let $j^M$ be the empirical current, defined as
in~\cite{Bertini2015}. Then, as $M\to\infty$, one has an LDP
\begin{equation}
 \mathrm{Prob}\left((\rho^M_t,j^M_t)_{t\in[0,\tau]} \approx (\rho_t,j_t)_{t\in[0,\tau]} \right) 
 \asymp \exp\left( - M I_{[0,T]}(\rho,j) \right) ;
 \label{equ:over-LDP}
\end{equation}
The rate function $I_{[0,T]}$ is finite only if $\partial_t\rho = - \nabla\cdot j$, in which case
\begin{equation}
 I_{[0,T]}(\rho,j) = \frac{1}{T} \Vc(\rho_0) + \frac{1}{4T} \int_0^\tau 
 \left\langle j_t-J(\rho_t),\, \chi(\rho_t)^{-1} \big(j_t-J(\rho_t)\big) \right\rangle \mathrm{d}t , 
 \label{equ:over-LDP-rate}
\end{equation}
where $\Vc(\rho)=T\int_\Gamma \rho(x) [\log \rho(x) + U(x) + \log Z ] \dd x$ is the quasipotential
(a kind of non-equilibrium free energy) and
\begin{equation} 
 J(\rho)=\chi(\rho)F(\rho), \qquad \chi(\rho)=\rho\hat\gamma, \qquad F(\rho)= f - T\nabla\log\rho. 
 \label{equ:J-chi-F-over} 
\end{equation}
Physically, $\chi$ is a mobility and $F$ is a force that acts in the space of densities (distinct from the physical force $f$).

The adjoint process obeys an LDP that is analogous to~\eqref{equ:over-LDP} and \eqref{equ:over-LDP-rate}, with $J(\rho)$ replaced by
$J^*(\rho)=\chi(\rho)F^*(\rho)$, where the adjoint force $F^*$ can be obtained from the following formulae, \linebreak which
mirror~\eqref{equ:fs-fa-over}:
\begin{equation}
 F^S(\rho) = \frac{F(\rho)+F^*(\rho)}{2} = f^S - T\nabla \log \rho, \qquad F^A = \frac{F(\rho)-F^*(\rho)}{2} = f^A .
\end{equation}

The resulting theory has several interesting features. First, within this general framework~\cite{Bertini2015}, the force $F^S$
is a free energy gradient, and is orthogonal to $F^A$ in the sense that
\begin{equation}
 F^S = -\nabla \frac{\delta\Vc}{\delta \rho}, \qquad 
 \int_\Gamma F^A \cdot \chi F^S \, \mathrm{d}x = 0 .
 \label{equ:FS-FA-over}
\end{equation}
This also implies that $\langle J^A,F^S\rangle=0=\langle J^S,F^A\rangle$. Second, we have an LDP analogue
of~\eqref{equ:Q-hk-diff} and~\eqref{equ:Qhk-fA-diff}, which follows directly from~\eqref{equ:over-LDP} and reads
\begin{equation}
 {\cal Q}^{\rm hk}(\rho,j) := 
 \lim_{M\to\infty}
 \frac1M \log 
 \frac{\mathrm{Prob} \left((\rho^M,j^M)_{t\in[0,\tau]} \approx (\rho,j)_{t\in[0,\tau]} \right)}
 {\mathrm{Prob}\left((\rho^M,j^M)_{t\in[0,\tau]} \approx (\rho_{\tau-t},-j_{\tau -t})_{t\in[0,\tau]}\right)} 
 =
 \int_0^\tau \langle j_t , F^A \rangle \mathrm{d}t.
 \label{equ:QHK-LDP-over}
\end{equation}
Note that $Q^{\rm hk}$ in~\eqref{equ:Q-hk-diff} is the heat transfer for a given sample path: here we are defining
${\cal Q}^{\rm hk}$ as the average heat transfer for a family of paths, as specified by $\rho$ and $j$. The antisymmetric force
$F^A$ is responsible for the housekeeping heat. In the steady state one has $\phi=\phi_U:=(\ee^{-U}/Z)$ and the associated
empirical current is $j^U=J^A(\phi_U)$; in this case ${\cal Q}^{\rm hk}=\tau \langle J^A(\phi_U),F^A \rangle$ depends only on the
anti-symmetric force and current.

\subsubsection{Physical Significance and Relation to Molecular Dynamics}
\label{sec:over-phys}

The key results from this section are (i) that splitting the physical force $f=f^S+f^A$ establishes a connection between $f^A$ and
the housekeeping heat (which determines the steady-state dissipation)~\cite{Seifert2012rmp}; (ii) that splitting the probability
current $J=J^S+J^A$ shows that $J^S$ corresponds to a gradient flow for the quasipotential $\Vc$, within an appropriate
metric~\cite{Bertini2015,adams13,jack-zimmer14}; and (iii) that the currents $J^S$ and $J^A$ are orthogonal, which allows the
characterisation of the quasipotential as the solution of a Hamilton-Jacobi equation~\cite{Bertini2015} and also has consequences
for the rate of convergence of such systems to their steady states~\cite{kaiser17accel}. We recently showed
in~\cite{kaiser17-lev2.5} that these structures are also present in (irreversible) Markov chains, although the notion of
orthogonality needs to be generalised, and the rate function analogous to~\eqref{equ:over-LDP-rate} is not a quadratic function of
the current in that case.

{From a physical point of view, the decomposition of the force as $f^S+f^A$ means for any (irreversible, non-equilibrium) diffusion process, one can define a reversible process in which the force $f^S$ acts alone, and this process has the same invariant measure as the original one (where $f=f^S+f^A$). If one considers many copies of this system as in Section~\ref{sec:ldp-over} then the reversible process evolves by steepest descent of the free energy, while the non-conservative component of the dynamics ($f^A$) gives rise to a probability current $J^A$ that flows in a direction orthogonal to the free-energy gradient. The reversible sector of the theory includes all information about the invariant measure [via~\eqref{equ:FS-FA-over}], while the irreversible sector describes the entropy production, as shown by~\eqref{equ:Qhk-fA-diff} and~\eqref{equ:QHK-LDP-over}. The orthogonality of the forces in~\eqref{equ:FS-FA-over} ensures that the decomposition of the force is unique, although obtaining explicit formulae for $f^S$ and $f^A$ requires that the invariant measure of the system is known, which is not typically the case for irreversible processes. As an analogy for the splitting, one~may think in terms of a Helmholtz decomposition of the force into a gradient ($f^S$) and a circulation ($f^A$), or perhaps as a functional Hodge decomposition of the probability current into three pieces, as~in~\cite{decarlo17} (see also~\cite{kaiser17accel}). Regardless of the specific mathematical structure, the key point is that we obtain a decomposition of the forces and currents into two parts, with distinct geometrical properties, and different physical~interpretations.

If we return briefly to the example of Section~\ref{sec:eg} and Figure~\ref{fig:eg} and consider the overdamped limit (with $f^{\rm ext}>0)$, one expects the following properties. The qualitative features of the potential $U$ will be given by the negative of the logarithm of the ``non-equilibrium'' distribution shown in Figure~\ref{fig:eg}b: it~will have a single minimum at some $q>0$. The force $f^S$ is simply the gradient of this potential, and~the reversible process in which $f^S$ acts alone is simply a diffusion in this potential. The non-reversible force is not the gradient of a potential: it is positive on average, so that it drives the system around the circle. However, it is not a constant force like $f^{\rm ext}$, it has a non-trivial dependence on the co-ordinate $q$, so that the physical force $f=f^{\rm ext}-V'$ is recovered as $f^S+f^A$.}

{We emphasise, however, that the results presented so far in this section are restricted to overdamped dynamics, and follow directly from macroscopic fluctuation theory~\cite{Bertini2015}. Our aim now is to extend them}
to molecular dynamics, as given by~\eqref{equ:ham-eom}
and~\eqref{equ:noneq-eom}. We will show that there are two possible extensions of the overdamped case, which corresponds to two
different splittings of the current $J$. One of the choices yields a geometrical structure analogous to~\eqref{equ:FS-FA-over},
which is related to the GENERIC (General Equation for Non-Equilibrium Reversible-Irreversible Coupling)
formalism~\cite{ott97a,ott97b}: see~\cite{duong13,kraaij17}. However, there is no connection between this splitting and the
housekeeping heat. The~second splitting makes the connection to the housekeeping heat, similar to~\eqref{equ:QHK-LDP-over}, but
there is no gradient structure analogous to~\eqref{equ:FS-FA-over}. We briefly discuss the advantages and disadvantages
of the two approaches: {their~physical consequences are addressed in Section~\ref{sec:split-outlook}}.

\subsection{Extension to Systems with Finite Damping: (pre)-GENERIC Splitting}
\label{sec:split-generic} 

We consider the model of~\eqref{equ:noneq-eom}, which we write as
\begin{equation}
 \dd q_t = p_t \dd t,
 \qquad
 \dd p_t = \gamma \Fc_p(q_t) \dd t + \sqrt{2\gamma T} \dd W_t , 
 \label{equ:noneq-eom-Fc}
\end{equation}
with a (rescaled) force $\Fc_p = (f/\gamma) - p$. The analysis of this section follows closely that of Section~\ref{sec:over}, with
the state point $x$ replaced by the phase space point $(q,p)$. Note, however, that there is no noise in the equation of motion
for the co-ordinates $q^i$, so some of the friction constants $\gamma_i$ in~\eqref{equ:sde-over} must be set to zero. The
implications of this will be discussed below. The Fokker-Planck equation for this system involves a phase space density $\phi$
defined on the space $\Omega$: we write
\begin{equation}
 \partial_t \phi = - \nabla \cdot J(\phi)
 \label{equ:fp-ham}
\end{equation}
with $J=(J_q,J_p)$ and $\nabla=(\nabla_q,\nabla_p)$ having components for both co-ordinates and momenta. Specifically,
\begin{equation} 
 J_q(\phi) = \phi p , \qquad J_p(\phi)=\gamma\phi\Fc_p - \gamma T \nabla_p \phi . 
 \label{equ:Jp-Jq} 
\end{equation}
The invariant measure is $\pi$ and we write
\begin{equation} 
 \dd\pi(q,p) = \frac{\ee^{-U(q,p)}}{ Z } \dd(q,p), 
 \label{equ:pi-U} 
\end{equation}
by analogy with~\eqref{equ:inv-over}. In this case, $U$ may be obtained by solving~\eqref{equ:UU}.

\subsubsection{Adjoint Process}

We define the adjoint process exactly as in~\eqref{equ:P*}. Note that this definition does not involve the reversal of any
momenta. The construction of the adjoint is given in Appendix~\ref{app:generic}. Its equation of motion involves the adjoint
force $\Fc_p^*$:
\vspace{-6pt}
\begin{equation}
 \dd q_t = -p_t \dd t,
 \qquad
 \dd p_t = \gamma \Fc^*_p(q_t) \dd t + \sqrt{2\gamma T} \dd W_t . 
\end{equation}
Note that the rate of change of $q$ is now in the \emph{{opposite}} direction to $p$, because the operator $\TT_0$ reverses time
without flipping the momenta. The analogue of~\eqref{equ:fs-fa-over} is
\begin{equation}
 \Fc^S_p = \frac12 (\Fc_p + \Fc_p^*) = -T \nabla_p U, \qquad \Fc^A_p = \frac12 (\Fc_p - \Fc_p^*) = (f/\gamma) - ( p - T\nabla_p U).
 \label{equ:def-ws-wa}
\end{equation}

For the case of conservative forces as in~\eqref{equ:ham-eom}, one has $f=-\nabla V$ for some potential $V$, so that
$U=(p^2/2+V)/T$. In this case, the antisymmetric force $\Fc^A_p$ contains the Hamiltonian evolution, while the symmetric force
contains the coupling to the thermostat. That is the essence of the GENERIC formalism~\cite{ott97a,ott97b}: see
also~\cite{duong13,kraaij17}. This connection is clearer at the level of probability currents, as we discuss in the next section.
However, in contrast to the overdamped case, we note that splitting the force as $\Fc_p=\Fc_p^S+\Fc_p^A$ does not provide a
general connection to dissipation: for these systems, the~heat flow is given by~\eqref{equ:Q-PX-PTX}, which breaks the analogy
with the overdamped case, where the formula~\eqref{equ:Q-PX-diff} applies. It is not possible to apply~\eqref{equ:Q-PX-diff} in
systems with finite damping, because the adjoint process has $\partial_t q = -p$, so $\dd \Pc(X)>0$ implies $\dd \Pc^*(X)=0$
(unless by some chance $p_t=0$ for all $t$).

\subsubsection{Large Deviation Principle}

We now analyse large deviations in these systems, following a method that is parallel to Section~\ref{sec:ldp-over}. We consider $M$
copies of our system, and let $(\phi^M,j^M)$ be the empirical density and current, defined on the phase space $\Omega$. The
analogue of~\eqref{equ:over-LDP} is
\begin{equation}
 \mathrm{Prob}\left((\phi^M,j^M)_{t\in[0,\tau]} \approx (\phi,j)_{t\in[0,\tau]} \right) \asymp \exp\left( -M I_{[0,T]}(\phi,j) \right). 
 \label{equ:ham-LDP}
\end{equation}

We write $j=(j^q,j^p)$, and the rate function is finite only if $j^q=J_q=p\phi$ [recall~\eqref{equ:Jp-Jq}] and
$\partial_t\rho = -\nabla\cdot j$, in which case
\begin{equation}
 I_{[0,T]}(\rho,j) = 
 \frac{1}{T} \Vc(\phi_0) -\frac{1}{4T} \int_0^\tau 
 \left\langle j_t^p-J_p(\phi_t),\, \frac{1}{\chi_p(\phi_t)} \big(j_t^p-J_p(\phi_t)\big) \right\rangle \mathrm{d}t ,
 \label{equ:ham-LDP-rate}
\end{equation}
with $\Vc(\phi)=T \int_\Omega \phi(q,p) [\log \phi(q,p) + U(q,p) + \log Z] \dd(q,p)$, also $\chi_p(\phi) = \gamma\phi$, and $J_p$
was defined in~\eqref{equ:Jp-Jq}. For a complete analogy with Section~\ref{sec:ldp-over}, one should take a mobility matrix
$\chi=\mathrm{diag}(\chi_q,\chi_p)$ with $\chi_q=0$: however, the fact that $\chi$ is singular means that not all results from the
overdamped case can be applied in this setting: see below.

We seek an analogue of~\eqref{equ:FS-FA-over}. That is, our aim is to split the current $J$ into two orthogonal components, one of
which is a free-energy gradient. To this end, we consider an LDP for the adjoint process, which is analogous
to~\eqref{equ:ham-LDP} and~\eqref{equ:ham-LDP-rate}, but with $J_p$ replaced by
$J_p^*(\phi)=\chi_p\Fc^*_p - \gamma T \nabla_p \phi$ and with the modified constraint that $j^q=J^*_q=-p\phi$ (instead of
$+p\phi$). Hence defining $J^S=(J+J^*)/2$, and~$J^A=(J-J^*)/2$, one has
\begin{align}
 J_p^S &= - \gamma T (\phi\nabla_p U + \nabla_p \phi), & J_p^A & = \phi \bigl[ f(q) - \gamma( p - T \nabla_p U) \bigr],
                                  \nonumber\\
 J_q^S &= 0, & J_q^A & = p\phi.
 \label{equ:JS-JA-ham-explicit}
\end{align}

Since $\chi$ is a singular matrix, it is not possible to define forces $F$ such that $J(\phi) = \chi(\phi) F(\phi)$,
 in~contrast
to~\eqref{equ:J-chi-F-over} for the overdamped case. However, since $J^S_q=0$, it is possible to write
\begin{equation}
 J^S(\phi) = \chi(\phi) F^S(\phi), \qquad F^S= - \nabla \frac{\delta\Vc}{\delta\phi} .
 \label{equ:FS-grad-generic}
\end{equation}
This allows us to identify the symmetric part of the dynamics as a gradient flow. Moreover, by~direct analogy
with~\eqref{equ:FS-FA-over}, it may be verified that
\begin{equation}
 \left\langle J^A, F^S \right\rangle = \int_\Omega \big[ J^A(\phi) \cdot F^S(\phi) \big] \dd(q,p) = 0 , 
 \label{equ:orth-generic}
\end{equation}
which says that the antisymmetric current is orthogonal to the gradient of the quasipotential. Using~\eqref{equ:FS-grad-generic}
and integrating once by parts, it follows that the quasipotential is constant under the antisymmetric part of the time evolution,
see also~\cite{kaiser17accel}.

As noted above, in the conservative case, where $f=-\nabla_q V$, then $U=H/T$ with $H=(p^2/2)+V$, and so $T\nabla_p U=p$. In that
case the anti-symmetric current contains the terms coming from the Hamiltonian evolution: $J^A=(\nabla_p H,-\nabla_q H)\phi$ and
$J^S$ contains the terms proportional to $\gamma$, which~come from the coupling to the heat bath. This is the setting that has
been named pre-GENERIC~\cite{kraaij17}. However, we emphasise that~\eqref{equ:FS-grad-generic}
and~\eqref{equ:orth-generic} apply also in the non-conservative setting. We also note that the absence of noise in the equation
of motion for $q$ makes $\chi$ singular: it is possible to regularise this system by adding an independent noise that acts on $q$,
which does not change any of the conclusions of this section.

The geometrical structure that is apparent from~\eqref{equ:FS-grad-generic} and~\eqref{equ:orth-generic} makes the construction of
this section attractive. One can view a general time-evolution as superposition of a gradient flow towards the non-equilibrium
steady state, together with an orthogonal drift that breaks time-reversal. However, the overdamped case also includes formulae
such as~\eqref{equ:Q-hk-diff} and~\eqref{equ:QHK-LDP-over}, which relate the antisymmetric forces and currents to dissipation. As
noted above, these results have no analogues in this setting: the~connection between the splitting of $J$ and the dissipation has
been lost in the passage from overdamped systems to those considered here. For this reason, we now consider an alternative
splitting of the current $J(\phi)$ that appears in~\eqref{equ:ham-LDP}. This alternative splitting loses the gradient structure
encoded in~\eqref{equ:FS-grad-generic}, but recovers the connection to the heat flow.

\subsection{Splitting the Currents and Forces into Equilibrium and non-Equilibrium Components}
\label{sec:split-dual}
\vspace{-6pt}
\subsubsection{Dual Process}

{We introduce a dual process, which differs from the adjoint process defined above. The~nomenclature ``dual process'' is discussed
in Appendix~\ref{app:dual}.} (The idea of comparing path measures for different processes in order to make connections to heat
flow is discussed in~\cite{Seifert2012rmp}.) The path measure for the dual process is $\Pcbar$. It obeys
\begin{equation}
 \dd\Pcbar(X)=\dd\Pc(\TT X) , 
 \label{equ:Pbar}
\end{equation}
which differs from~\eqref{equ:P*} since the operator $\TT$ reverses all momenta (recall Section~\ref{sec:heat-time}).
Applying~\eqref{equ:Pbar} for paths with $\tau\to0$, one sees that the invariant measure $\pibar$ of the dual process satisfies
$\dd\pibar(q,p) = \dd\pi(q,-p)$, so the analogue of~\eqref{equ:pi-U} is
\begin{equation}
 \dd\pibar(q,p) = \frac{ \ee^{-\Ubar(q,p)} }{Z } \dd (q,p), \qquad \Ubar(q,p)=U(q,-p).
 \label{equ:pibar}
\end{equation} 

The dual process may be constructed: see Appendix~\ref{app:dual}. The coordinates and momenta in the dual process obey
\begin{equation}
 \dd q_t = p_t \dd t ,  \qquad \dd p_t = f(q_t) \dd t - \gamma\left(2T \nabla_p\Ubar - p_t \right) \dd t + \sqrt{2\gamma T} \dd W_t . 
 \label{equ:sde-adj}
\end{equation}
As above, we write the deterministic term in the equation of motion for $p$ (in the original process) as $\gamma \Fc_p \dd t$ with
$\Fc_p = (f/\gamma)-p$. The corresponding quantity in the dual process is
\begin{equation}
 \Fcbar_p = (f/\gamma)+p-2T \nabla_p \Ubar . \label{equ:wbar} 
\end{equation} 
In the conservative case, $\Ubar=H/T$, so $T\nabla_p\Ubar=p$ and the dual process coincides with the original process. Since the
conservative case corresponds to a model with an equilibrium steady state, we~define
\begin{equation}
 \Fc^E_p := \frac12(\Fc_p+\Fcbar_p) = (f/\gamma) -T \nabla_p \Ubar , \qquad 
 \Fc^N_p := \frac12(\Fc_p-\Fcbar_p) = T \nabla_p \Ubar - p , 
 \label{equ:fep-fen}
\end{equation}
where the superscripts $E$ and $N$ indicate equilibrium and non-equilibrium contributions. These are the analogues of the
symmetric and antisymmetric forces discussed above.

\subsubsection{Formulae for Heat Currents Based on Sample Paths}

The housekeeping heat for these systems is defined by analogy with~\eqref{equ:Q-hk-diff} as 
\begin{equation}
 Q^{\rm hk}(0,\tau) = T \log \frac{\dd\Pc_{X_0}(X)}{\dd\Pcbar_{X_0}(X)}.
 \label{equ:Q-hk-ham}
\end{equation}
Note that the path probabilities in~\eqref{equ:Q-hk-ham} are conditioned on their initial states as in~\eqref{equ:Q-hk-diff}. This
is essential, so as to ensure that $Q^{\rm hk}(0,\tau)\to0$ as $\tau\to0$: as the trajectory length goes to zero, so does the heat
flow. Recalling~\eqref{equ:Q-PX-PTX}, the analogue of~\eqref{equ:QQ-diff} is that for any path $X$
\begin{equation}
 Q(0,\tau) = Q^{\rm hk}(0,\tau)  + T \log \frac{\dd\pi((\TT X)_0)}{\dd\pibar(X_0)} = Q^{\rm hk}(0,\tau) 
 - T \int_0^\tau \dd\Ubar_t , 
\end{equation}
where the second equality uses~\eqref{equ:pi-U},~\eqref{equ:pibar} and $\Ubar(q,-p)=U(q,p)$. Using~\eqref{equ:dE-not-dH} to
substitute for $Q$, one~sees that
\begin{equation}
 Q^{\rm hk}(0,\tau) = \int (T \nabla_q \Ubar+f) \circ \dd q_t + (T \nabla_p \Ubar - p_t)\circ \dd p_t. 
 \label{equ:qhk-fn-prelim}
\end{equation}

The analogy with~\eqref{equ:Qhk-fA-diff} motivates us to define a ``force'' acting in the phase space as 
\begin{equation} 
 \Fc^N = (\Fc^N_q,\Fc^N_p)=(T\nabla_q \Ubar+f, T\nabla_p \Ubar-p). 
 \label{equ:Fn-full} 
\end{equation}
This is a non-equilibrium force, in the sense that it vanishes in conservative systems. (Recall that the conservative case has
$U=(p^2/2 + V)/T$ and $f=-\nabla_q V$.) Hence, in terms of dissipation, $\Fc^N$ is analogous to the force $f^A$ in the overdamped
case, and one has
\begin{equation}
 Q^{\rm hk}(0,\tau) = \int \Fc^N \circ (\dd q_t,\dd p_t) ,
 \label{equ:qhk-fn-final} 
\end{equation}
which shows that the ``non-equilibrium'' force $\Fc^N$ does indeed determine the steady-state dissipation.

\subsubsection{Large Deviation Principles}

These results also have implications for large deviations. For the original process one still has~\eqref{equ:ham-LDP}. One
splits $J_p = J^E_p + J^N_p$ such that the corresponding LDP for the dual process is similar, but now with
$\Jbar_p = J^E_p - J^N_p$. In this case
\begin{align}
 J^E_p(\phi) & = \phi\gamma \Fc_p^E - \gamma T \nabla\phi, & J^N_p(\phi) & = \phi \gamma \Fc^N_p , 
                                 \nonumber       \\                           
 J^E_q(\phi) & = \phi p, & J^N_q(\phi) & = 0.
 \label{equ:JE-JN-explicit}
\end{align}
One has $\chi_p=\phi\gamma $ and $\chi_q=0$, as in Section~\ref{sec:split-generic}. Since $\chi$ is singular, it is not possible to
write $J(\phi)=\chi(\phi) F(\phi)$, but one does have
\begin{equation}
 J^N(\phi) = \chi(\phi) \Fc^N .
\end{equation}

Moreover, there is an orthogonality relation analogous to~\eqref{equ:orth-generic}. This is derived in Appendix~\ref{app:dual}.
The result is that
\begin{equation}
 \left\langle J^E, \Fc^N \right\rangle = \int_\Omega \big[ J^E (\phi) \cdot \Fc^N \big] \dd(q,p) = 0.
 \label{equ:orth-FN-JE}
\end{equation}
It is clear that $J^E$ does not correspond to a gradient flow, so there is no analogue of~\eqref{equ:FS-grad-generic}. However,
there is an analogous statement to~\eqref{equ:QHK-LDP-over}, which is that the housekeeping heat flow associated with the path
$(\rho,j)$ is
\begin{equation}
 {\cal Q}^{\rm hk}(0,\tau) 
 = \int_0^\tau \left\langle j_t , \Fc^N \right\rangle \mathrm{d}t .
 \label{equ:QHK-LDP-ham}
\end{equation}
The steady state probability density is $\phi=\phi_U$ and the associated empirical current is \linebreak$j^U=J^E(\phi_U)+J^N(\phi_U)$, with
both ``equilibrium'' and ``non-equilibrium'' currents contributing, contrary~to the overdamped case. However,
from~\eqref{equ:orth-FN-JE} one has ${\cal Q}^{\rm hk}=\tau \langle J^N(\phi_U),F^N \rangle$, which depends only on the
non-equilibrium force and current. Thus the non-equilibrium part of the theory is intrinsically linked to the housekeeping heat,
as one might expect from the definitions~\eqref{equ:Pbar} and~\eqref{equ:Q-hk-ham}.

\subsection{Discussion}
\label{sec:split-outlook}

In Section~\ref{sec:over}, we reviewed some results that show how dynamical fluctuations in overdamped systems are accompanied by
underlying geometrical structures related to gradient flows, orthogonalities and dissipation. In Sections~\ref{sec:split-generic}
and~\ref{sec:split-dual}, we showed how these structures generalise to systems described by Hamiltonian dynamics, including
non-equilibrium driving forces, and coupling to a heat bath. The resulting structures are more complex, since there are two
alternative time-reversal operations, depending on whether one chooses to reverse the momenta or not.

To summarise the key results: one may split the probability current either as \mbox{$J=J^S+J^A$} (corresponding to simple time-reversal)
or as $J=J^E+J^N$ (corresponding to time-and momentum-reversal). In both cases, the resulting currents (and their conjugate
forces) are orthogonal, in the sense of~\eqref{equ:orth-generic} and~\eqref{equ:orth-FN-JE}. 
It is likely that these orthogonalities can be used to derive bounds on the rates with which non-equilibrium systems converge to their steady states,
by generalising the analysis of~\cite{Spiliopoulos2016,kaiser17accel,kaiser17-lev2.5}.

The splitting $J=J^S+J^A$ recovers the recently proposed
(pre-)GENERIC splitting of~\cite{kraaij17}, at least for conservative systems. In this case, two currents $J^S$ and $J^A$ can be identified straightforwardly,
since $J^A$ corresponds to the Hamiltonian evolution and the $J^S$ to the action of the thermostat. 
This~decomposition is very natural in that context, and 
is exploited (for example) in integration schemes for molecular dynamics~\cite{leimBook} (in the ``{BAOAB}'' 
notation of that work, $J^A$ encapsulates the parts of the evolution denoted by A,B and $J^S$ is responsible for the part denoted by O). The fact that this same decomposition of the current can be applied in non-conservative systems is not so well-known---this case resembles the overdamped situation of Section~\ref{sec:over}, where the two parts of the current have the same geometrical properties, as steepest descent of the free energy ($J^S$), and an orthogonal drift ($J^A$). The properties of $J^S$ connect this splitting to recent studies that represent convergence to a steady state as a gradient flow for the free energy~\cite{adams13,mielke14}. However, as in the overdamped case, explicit formulae for $J^S$ and $J^A$ are not available, and computing these quantities is only possible if the invariant measure (or quasipotential) of the system is known. In addition, the current $J^A$ is not connected to the entropy production in this case---in this sense, the splitting does not separate the different aspects of the system as cleanly as was the case for overdamped systems.

On the other hand, if one considers the splitting $J=J^E+J^N$ then there is no sense of steepest descent of the free energy ($J^E$ is not a gradient), but this splitting does provide a connection to the steady-state dissipation (via~\eqref{equ:QHK-LDP-ham}). In the absence of a gradient structure, the splitting does not provide as simple a {physical picture as in the overdamped case, but it is interesting to note that one may represent the time evolution of such a system as a combination of a non-dissipative process (described by $J^E$)} and a dissipating one $J^N$.

To illustrate these points, we return to the example of Section~\ref{sec:eg} and Figure~\ref{fig:eg}: if $f^{\rm ext}=0$ then the potential $U(q,p)=[p^2/2 + V(q)]/T$ is symmetric in both its arguments, with a single minimum at $(q,p)=0$. For $f^{\rm ext}>0$ then $U$ cannot be separated as a sum of terms depending on $q$ and $p$ alone, so the co-ordinates and momenta are not independent. Moreover, $U$ does not in general have any symmetry. One does expect a single minimum for some $q,p>0$. For the splitting $J=J^S+J^A$, we~note from~\eqref{equ:JS-JA-ham-explicit} that the phase space current $J^S$ acts only on the momentum co-ordinates: it represents the action of a thermostat that applies damping and noise, and drives the momentum distribution $P(p)=\int \phi(q,p) \mathrm{d}q$ towards its ($q$-dependent) ``local equilibrium'' form $P_{\rm eq}(p|q)=(1/Z) \int {\rm e}^{-U(q,p)} \mathrm{d}q$. The antisymmetric part of the dynamics, described by $J^A$, includes an (irreversible) advection of the co-ordinates in accordance with the current value of the local momentum, as well as the effect of the non-equilibrium forces $f$ on the momenta. In summary, one can think of $J^S$ as the result of a ``non-equilibrium thermostat'' in which the damping force $\mathcal{F}^S_p=-T\nabla_p U$ depends on the co-ordinates $q$ as well as the momenta $p$, and which drives the system towards a state with finite average momentum. The current $J^A$ accounts for the Hamiltonian parts of the time evolution, and the non-conservative~forces.

For the splitting $J=J^E+J^N$, the potential that appears is $\Ubar(q,p)=U(q,-p)$: one expects that this function has a minimum for some $q>0$ and $p<0$. The non-equilibrium current $J^N$ acts only on the momentum and can be interpreted in terms of a coordinate-dependent damping force, ${\cal F}^{N}_p(q,p)=-T\nabla_p [p^2/(2T)-\overline{U}(q,p)]$ which we expect (on average) to drive the system towards positive momenta. The current $J^E$ includes the advection of the co-ordinates by the momenta, as well as the action of the force $f$. For this current acting alone, one arrives at a process whose steady state is time-reversal-symmetric (in the sense that the right hand side of~\eqref{equ:Q-hk-ham} vanishes), but whose invariant measure is not provided by the above analysis (and is neither ${\rm e}^{-U}$ nor ${\rm e}^{-\Ubar}$). The nature of the process described by $J^E$ acting alone seems to deserve further investigation (both in this specific case and more generally).

As a final point, we note that the rate functions for path probabilities~\eqref{equ:over-LDP-rate} and~\eqref{equ:ham-LDP-rate}
were derived by considering many copies of our system, but the same formulae also govern large deviations at level
2.5~\cite{maesEPL2008,Bertini2014,Bertini2015empirical,chetrite2015control,kaiser17-lev2.5}. These LDPs involve rare events where
an unusual density or current is sustained over a long time period (in a single system). Such LDPs are closely related to level-1
LDPs such as~\eqref{equ:j-ldp-lev1}. Moreover, orthogonality formulae such as~\eqref{equ:FS-FA-over} allow rate functions at
level 2.5 to be decomposed into contributions that come from currents that are symmetric and antisymmetric under time-reversal,
with implications for the rate of convergence of non-equilibrium systems to their steady
states~\cite{Spiliopoulos2016,kaiser17accel}. Such decompositions have also been connected to recent results related to bounds on
dissipation in non-equilibrium steady states~\cite{gingrich16,pietzonka16,kaiser17-lev2.5}.

For extensions to this work, it is possible to combine the two splittings presented here, in order to split the current $J$ into
four pieces, which are separated according to their behaviour under the two operations $\TT$ and $\TT_0$. It is also of interest
to relax the restriction that there is no noise in the equation of motion for $q$. We hope to return to the resulting geometrical
structures in a later work.

\section{Conclusions}

We end with a few remarks as to the relevance of these results for modelling systems by molecular dynamics. Throughout this
article, we have focussed on general results such as symmetries and geometrical structures. For example, we showed in
Section~\ref{sec:pt} that ensembles conditioned on atypical currents retain a $\PP\TT$ symmetry that is not present in typical
non-equilibrium states~\cite{JackEvans}. Hence the conditioned ensembles seem to be in a different class from non-equilibrium
steady states.

In the analysis of Section~\ref{sec:orth}, we showed how currents and forces in molecular systems can be split in different ways,
based on the theories of stochastic thermodynamics~\cite{Seifert2012rmp} and MFT~\cite{Bertini2015}. We~believe that these
results are relevant for two reasons. First, the existence of gradient structures such as~\eqref{equ:FS-grad-generic} has
potential mathematical applications, in rigorous derivations of effective theories that apply on large length and time
scales~\cite{Fathi2016a}. The idea is that if a system evolves by steepest descent of some free energy, then any coarse-grained
description of that system should also be represented as a steepest decent (of the coarse-grained free energy). Second, the use of
orthogonality relationships to decompose currents (and their corresponding rate functions~\cite{kaiser17-lev2.5}) has the
potential to establish new constraints on fluctuations in non-equilibrium systems. We also note in passing that by identifying
currents and their conjugate forces, one may also decompose rate functions using a canonical
structure~\cite{maesEPL2008,maes08beyond}, which makes explicit the connections between antisymmetry under time-reversal and
fluctuation theorems~\cite{maes1999gibbs}).
We look forward to more work in these directions, and~their application in practical~contexts.

\vspace{6pt} 

\acknowledgments{{Robert Jack} thanks Mike Evans for many useful discussions about the material of Section~\ref{sec:pt}. Marcus Kaiser is supported by
 a scholarship from the EPSRC Centre for Doctoral Training in Statistical Applied Mathematics at Bath (SAMBa), under the project
 EP/L015684/1. Johannes Zimmer gratefully acknowledges funding by the EPSRC through project EP/K027743/1, the Leverhulme Trust (RPG-2013-261)
 and a Royal Society Wolfson Research Merit Award. }

%

\appendix

\section{Constructions of Adjoint Processes}

\subsection{Overdamped Case}
\label{sec:app-over}

Let $\WW={\cal L}^\dag$ be the adjoint of the generator associated with the process~\eqref{equ:sde-over}. (The operator $\WW$ is
also known as the Fokker-Planck operator or master operator). Let $\rho$ be a probability density on $\Gamma$. 
We~take
$\gamma_i=\gamma$ for compactness of notation, the general case is a straightforward extension. Equation~\eqref{equ:fp-over}
becomes
\begin{equation}
 \partial_t\rho = \WW \rho = - \gamma \nabla \cdot ( f \rho ) + \gamma T \nabla^2 \rho .
 \label{equ:W-over}
\end{equation}
The steady state has $\mathrm{d}\pi(x)\propto \ee^{-U(x)}\dd x$: hence $\WW \ee^{-U}=0$, so that
\begin{equation}
 -\nabla\cdot f + f\cdot \nabla U + T (\nabla U)^2 - T \nabla^2 U =0.
 \label{equ:U-over}
\end{equation}
The corresponding operator for the adjoint process is 
\begin{equation}
 \WW^* = \hat{\pi} \circ \WW^\dag \circ \hat{\pi}^{-1} , 
\label{equ:W*}
\end{equation}
where $\WW^\dag$ is the regular adjoint of $\WW$ (the Hermitian conjugate, or adjoint in $L^2$); also the operator $\hat\pi$ acts
as $\hat\pi g(x) = \ee^{-U(x)} g(x)$ and $\circ$ simply indicates the composition of operators: for two operators
$\hat{A},\hat{B}$, the notation $(\hat{A}\circ \hat{B})\phi$ simply means that $\hat A$ is applied to $\hat B\phi$. Since
$\ee^{U(x)}>0$ for all $x$, one~may identify $\WW^*$ as the adjoint of $\WW$ associated with the inner product
$\langle h,\rho\rangle_U := \int_\Gamma h(x) \ee^{U(x)} \rho(x) \mathrm{d}x$: that~is,
$\langle h,\WW\rho \rangle_U = \langle \WW^* h,\rho \rangle_U$.

To see that the operator $\WW^*$ generates the path measure~\eqref{equ:P*}, consider the original process started at $x_0$. Let
the probability measure for the state $x_t$ be $G_{x_0}^t$ and let $\dd G_{x_0}^t(y) = g_t(x_0,y) \mathrm{d}y$, so that
$g_t(x_0,\cdot)$ is a probability density. One may identify $g_t$ as a ``matrix element'' of the operator $\ee^{\WW t}$ by
writing $g_t(x,y)=(\ee^{\WW t})_{y,x}$, which can be defined via
$\int_\Gamma h(y) \ee^{\WW t}\rho(y) \mathrm{d}y = \int_{\Gamma} \int_\Gamma h(y) \cdot (\ee^{\WW t})_{y,x} \cdot \rho(x)
\mathrm{d}x \mathrm{d}y$, which must hold for all functions $h,\rho$ (in some suitable class).

All together, this means simply that $(\ee^{\WW t})_{y,x}$ is the probability (density) that the system ends at $y$, if it starts
at $x$ some time $t$ earlier. From~\eqref{equ:P*}, recalling that the initial conditions for $\Pc$ come from the invariant
measure $\pi$, and that the invariant measure of the adjoint process is also $\pi$, one sees that $\WW^*$ must satisfy
\begin{equation}
 (\ee^{\WW t})_{y,x} \cdot \ee^{-U(x)} = (\ee^{\WW^* t})_{x,y} \cdot \ee^{-U(y)}
 \label{equ:adj-matrix}
\end{equation}
for all $x,y\in\Gamma$ and all $t$. Combining this with the definition of the adjoint, which implies that
$(\WW^\dag)_{x,y}=\WW_{y,x}$, one arrives at~\eqref{equ:W*}.

To apply this equation, note that $\hat{\pi} \circ \nabla \circ \hat{\pi}^{-1} = (\partial U+\nabla)$, where we write
$\partial U=\nabla U$: with~this notation, the operator $\nabla$ always acts on all arguments to its right, while $\partial U$
simply indicates multiplication by a function. Noting also that $\nabla^\dag=-\nabla$, one has
\begin{equation}
 \WW^* \rho = \gamma f \cdot (\partial U+\nabla) \rho + \gamma T (\partial U+\nabla)^2 \rho . 
\end{equation}

Since $\rho$ is a probability density, conservation of total probability means that $\int \WW^*\rho \mathrm{d}x=1$ for any $\rho$:
this formula can be used to recover~\eqref{equ:U-over}. Hence, one finds
\begin{equation}
 \WW^* \rho = - \gamma \nabla \cdot ( f^* \rho ) + \gamma T \nabla^2 \rho , 
\end{equation}
where $f^* = -(2T\partial U+f)$ is the force that appears in the adjoint process: see~\eqref{equ:f*-over}. If $f=-\nabla V$ for
some potential $V$, then $U=V/T$ and $f^*=f$: this is the reversible case. Note that we have analysed the case where all the
$\gamma_i$ are equal, but the final result~\eqref{equ:f*-over} applies also in the general case.

\subsection{GENERIC Splitting}
\label{app:generic}

The construction of the adjoint process used in Section~\ref{sec:split-generic} follows exactly the method of
Appendix~\ref{sec:app-over}, replacing $X$ by $(q,p)$ and $\Gamma$ by $\Omega$. We also replace $\rho$ by the density $\phi$ on
phase space. The analogue of~\eqref{equ:W-over} is
\begin{equation}
 \label{equ:W-ham}
 \WW\phi = -p \cdot \nabla_q \phi - f \cdot \nabla_p \phi + \gamma \nabla_p \cdot ( p\phi) + \gamma T \nabla_p^2 \phi , 
\end{equation}
and the analogue of~\eqref{equ:U-over} is
\begin{equation}
 p\cdot \partial_q U +f \cdot \partial_p U + (\nabla_p - \partial_p U) \cdot \gamma ( p - T\partial_p U)=0.
 \label{equ:UU}
\end{equation}
Using~\eqref{equ:W*} to construct $\WW^*$, the equation of motion for the adjoint process may then be verified to
be~\eqref{equ:noneq-eom-Fc}.

\subsection{Dual Process}
\label{app:dual}
\vspace{-6pt}
\subsubsection{Construction of the Dual Process}

This section is analogous to Appendix~\ref{app:generic}, but is based on the dual process defined by~\eqref{equ:Pbar}. For
consistency with that definition, the operator $\WWbar$ that generates the dual process must satisfy an analogue
of~\eqref{equ:adj-matrix}, which is
\begin{equation}
 (\ee^{\WW t})_{y,x} \cdot \ee^{-U(x)} = (\ee^{\WWbar t})_{\TT x,\TT y} \cdot \ee^{-U(\TT y)} , 
 \label{equ:dual-matrix}
\end{equation}
where $x,y\in\Omega$ are phase space points: recall that the action of $\TT$ on phase space points is \mbox{$\TT(q,p)=(q,-p)$.} Define
an operator $\hat\TT$ that acts on functions as $\hat\TT g(q,p):=g(q,-p)$. With this definition,~\eqref{equ:dual-matrix} is
equivalent to
\begin{equation}
 \WWbar = \hat\TT \circ \hat{\pi} \circ \WW^\dag \circ \hat{\pi}^{-1} \circ \hat\TT . 
 \label{equ:Wbar}
\end{equation}
{We note that this is a duality mapping, since
$\WWbar \circ ({\hat\TT} \circ\hat{\pi}) = ({\hat\TT} \circ\hat{\pi}) \circ \WW^\dag$, but $\WWbar$ is not an adjoint of $\WW$.}
Hence our terminology ``dual process''. Also, $\hat\TT \circ U \circ \hat\TT = \Ubar$, and the analogue of~\eqref{equ:UU}
for the dual process is
\begin{equation}
 -p \cdot \partial_q \Ubar -f \cdot \partial_p \Ubar + (\nabla_p - \partial_p \Ubar) \cdot \gamma ( p - T\partial_p \Ubar)=0.
 \label{equ:UUbar}
\end{equation}
Hence, using~\eqref{equ:W-ham} and~\eqref{equ:pibar}, one has
\begin{equation}
 \WWbar \phi = -p \cdot \nabla_q \phi - \nabla_p \cdot [(f(q) - 2\gamma T\partial_p \overline{U} +\gamma p ) \phi] 
 + \gamma T \nabla_p^2 \phi , 
\end{equation}
from which one identifies the dual force $\fbar=f - 2\gamma T\partial_p \overline{U} +\gamma p$, consistent
with~\eqref{equ:sde-adj}.

\subsubsection{Orthogonality of Currents and Forces}

Here we derive the orthogonality formula~\eqref{equ:orth-FN-JE} of the main text. Defining $\Fc^N$ as in~\eqref{equ:Fn-full} and
$\Fc_p^E$ as in~\eqref{equ:fep-fen}, one has from~\eqref{equ:dPX-qp} and~\eqref{equ:Q-hk-ham} that
\begin{equation}
 Q^{\rm hk}(0,\tau) = \int \Fc_p^{N} \circ \dd p_t - \gamma \Fc^N_p\cdot \Fc_p^E \dd t - \gamma T \nabla_p \cdot\Fc_p^N \dd t . 
\end{equation}
Comparing with~\eqref{equ:qhk-fn-prelim} and using $\dd q = p\dd t$, one sees that
\begin{equation}
 \Fc_q^N \cdot p + \gamma \Fc_p^N \cdot \Fc_p^E + \gamma T \nabla_p \cdot \Fc_p = 0 , 
 \label{equ:fs-fa-orth-dual}
\end{equation}
which is analogous to the result~\eqref{equ:fs-fa-orth} in the overdamped case, and may be verified from~\eqref{equ:UUbar}. Now
write
\begin{align}
 \left\langle J^E, \Fc^N \right\rangle 
 = \int_\Omega (J^E \cdot \Fc^N) \dd(q,p) = \int [(\phi p\cdot \Fc^N_q ) 
 + (\phi\gamma \Fc_p^E \cdot \Fc_p^N) - (\Fc_p^N \cdot \gamma T \nabla_p \phi)] \dd(q,p) . 
\end{align}
Integrating by parts once and using~\eqref{equ:fs-fa-orth-dual}, the right hand side is zero and we
recover~\eqref{equ:orth-FN-JE}.

\bibliography{ent-thin}

\end{document}